\documentclass[12pt]{iopart}
\usepackage{graphicx}
\usepackage{iopams,amssymb,bm}

\begin{document}


\title{Wave fronts, pulses and wave trains in photoexcited superlattices behaving as excitable or oscillatory media} 
\author{J. I. Arana$^1$, L. L. Bonilla$^1$, H. T. Grahn$^2$}

\address{$^1$G.\ Mill\'an Institute of Fluid Dynamics, Nanoscience and Industrial
Mathematics, University Carlos III de Madrid, Av.\ Universidad 30, 28911 Legan\'es,
Spain\\
$^2$Paul Drude Institute for Solid State Electronics, Hausvogteiplatz 5--7, 10117 Berlin, Germany}
\ead{joseignacio.arana@uc3m.es}\ead{bonilla@ing.uc3m.es}\ead{htgrahn@pdi-berlin.de}

\date{\today}

\begin{abstract}
Undoped and strongly photoexcited semiconductor superlattices with field-dependent
recombination behave as excitable or oscillatory media with spatially discrete
nonlinear convection and diffusion. Infinitely long, dc-current-biased superlattices
behaving as excitable media exhibit wave fronts with increasing or decreasing profiles,
whose velocities can be calculated by means of asymptotic methods. These superlattices
can also support pulses of the electric field. Pulses moving downstream with the flux of
electrons can be constructed from their component wave fronts, whereas pulses advancing upstream do so slowly and experience saltatory motion: they change slowly in long intervals of time separated by fast transitions during which the pulses jump to
the previous superlattice period. Photoexcited superlattices can also behave as
oscillatory media and exhibit wave trains.
\end{abstract}


\maketitle

\section{Introduction}
\label{sec:intro}
Vertical nonlinear charge transport in undoped, weakly coupled, photoexcited
superlattices (SLs) has many similarities to extended excitable or oscillatory
media. In both cases, traveling waves such as wave fronts, pulses, or wave
trains play an important role in the dynamics of the system.\cite{Bon10}
In undoped SLs, the formation and dynamics of electric field domains (EFDs)
are the basic mechanisms behind more complex phenomena such as self-sustained
oscillations of the current (SSOC) through voltage-biased SLs,  chaos, etc.
Interfaces separating EFDs are wave fronts or wave front-bound pulses, and
pulses are the building blocks of wave trains. While there are at least two
stable EFDs with an almost constant value of the electric field in the case
of doped SLs, there may be only one stable EFD in the case of an undoped SL
under high photoexcitation.\cite{Ara10} This is similar to what
happens in excitable media such as those described by the FitzHugh-Nagumo
(FHN) model for nerve conduction.\cite{Fit61,Nag62} For background
information on this model and related models, see
Refs.~\cite{Sco75,Bel84,Neu97,Str97,Kee98,Mci99,Mur01,Bon10}. The case of an
oscillatory medium in which there is a single unstable spatially homogeneous
state can also be realized in both the FHN model and a strongly photoexcited
undoped SL.\cite{Ara10} However, the electric field in a voltage-biased SL
with a finite number of periods is constrained to have a fixed area underneath its
spatial profile, and this constraint causes it to exhibit a more complex
behavior.\cite{Ara10}

The key ingredient to the complex dynamics of electron transport in photoexcited
undoped SLs is the introduction of an electron-hole recombination coefficient,
which is a function of the electric field. Nonlinear charge transport in weakly
coupled undoped type-I superlattices (SL) under photoexcitation is well
described by spatially discrete drift-diffusion systems (DDDS) of
equations.\cite{Bon94,Bon95,Bon05} In earlier models, the electron-hole
recombination was considered to be a constant, independent of the electric
field.\cite{Bon94} In this case, there are three stable states with spatially
homogeneous field profiles, and the predicted nonlinear phenomena are then
quite similar to the ones observed in the much better known case of doped SLs
(in which the hole density is replaced by a constant doping density):
the dynamics is organized by stable wave fronts that join the stable
homogeneous states.\cite{Bon05} Introducing a field-dependent electron-hole
recombination coefficient in the model has striking consequences.\cite{Ara10}
For high photoexcitation densities, it is possible to find only one stable EFD, not
two as in the case of a constant recombination coefficient.\cite{Bon94}
Then, under a dc voltage bias, periodic or chaotic SSOC may appear. The field
profile during self-oscillations may exhibit nucleation of dipole waves
inside the sample, splitting of one wave in two, and motion of the resulting
waves in opposite directions. Despite the strong asymmetry induced by the
drift terms in the equations, these dipole waves resemble the pulses in
the FHN model for nerve conduction\cite{Kee98,Mur01} and are quite different
from field profiles for a field-independent recombination coefficient.\cite{Bon94}

In this paper, we construct relevant stable solutions of the DDDS of
equations for dc-current-biased SL by means of asymptotic and numerical
methods that extend and refine those used for the discrete
FHN system.\cite{Car05,Car03} For an undoped SL that acts as an excitable
medium (only one stable EFD), we find that electric field pulses moving
downstream with the electron flux can be described using matched asymptotic
expansions based on separating the sharp leading and trailing wave fronts
of the pulse from smoother regions outside them. Pulses moving upstream do
so much more slowly and experience a saltatory motion, in which intervals
of a slow change are separated by fast changes during which the pulse jumps
to the previous SL period. For an undoped SL acting as an oscillatory
medium (a single unstable EFD), we construct periodic wave trains consisting
of an infinite succession of pulses.

\section{Model equations}
\label{sec:model}
\begin{figure}[!b]
\centerline{\includegraphics[width=0.5\linewidth]{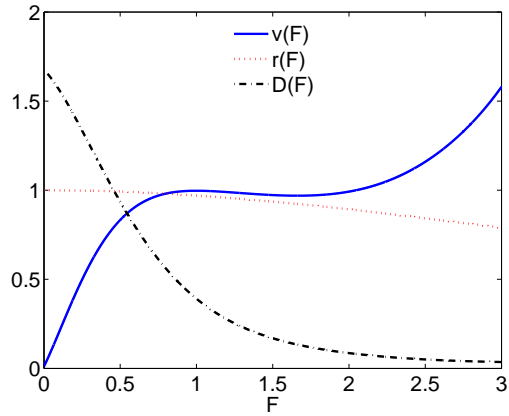}}
\caption{(Color online) Electron velocity $v$, diffusion coefficient $D$, and recombination
coefficient $r$ vs electric field $F$ for a GaAs/Al$_{0.3}$Ga$_{0.7}$As SL
with a well width $L_W=10$ nm and barrier width $L_B=4$ nm.
The lattice temperature is assumed to be 200~K.}
\label{fig1}
\end{figure}

Written in nondimensional form, the equations governing nonlinear charge
transport in a weakly coupled, undoped, photoexcited GaAs/Al$_x$Ga$_{1-x}$As SL
are:\cite{Ara10}
\begin{eqnarray}
&& F_{i}-F_{i-1}= \nu\, ( n_{i} -p_{i})\;,
\label{eq1}\\
&& \delta\, \frac{dF_{i}}{dt} + n_i v(F_i) - D(F_i)(n_{i+1}-n_i) = J\;,
\label{eq2}\\
&&\frac{dp_{i}}{dt}=1 - r(F_{i})\, n_{i}p_{i}\;,
\label{eq3}
\end{eqnarray}
where $i$ takes on integer values corresponding to the spatial periods of the
SL. $-F_{i}$, $n_{i}$ and $p_{i}$ are the average electric field, electron and
hole surface densities at the $i^{th}$ SL period, respectively, and the
dimensionless parameters $\nu$, $\delta$, and $r$ are described below.
Equation~(\ref{eq1}) refers to the averaged Poisson equation, while Eq.~(\ref{eq2})
corresponds to Amp\`ere's law: the total current density $J$ equals the sum of
the displacement current density $\delta\, dF_i/dt$ and the electron tunneling
current density across the $i^{th}$ barrier that separates wells $i$ and $i+1$,
i.e., $J_{i\to i+1}=n_i v(F_i) - D(F_i)(n_{i+1}-n_i)$. Note that the tunneling
current density has a discrete drift term with electron velocity $v(F)$ and
a diffusive term with diffusion coefficient $D(F)$. Charge continuity is obtained
by differentiating Eq.~(\ref{eq1}) with respect to time and using Eq.~(\ref{eq2})
in the result:
\begin{eqnarray}
\nu\delta\, \frac{d}{dt}( n_{i} -p_{i}) + J_{i\to i+1}-J_{i-1\to i} = 0\;.
\label{eq4}
\end{eqnarray}
Tunneling of holes is neglected so that only photogeneration and recombination
with electrons enter Eq.~(\ref{eq3}). The field-dependent quantities $v$,
$D$, and the recombination coefficient $r$ are described in Ref.~\cite{Ara10}
and their field dependencies are depicted in Fig.~\ref{fig1}. The dimensionless
parameter $\nu$ is a ratio between the carrier density scale determined by
photogeneration and recombination and the carrier density determined by
scattering processes. The dimensionless parameter $\delta$ is the
ratio between the time scales of dielectric relaxation and recombination.\cite{Ara10}
These parameters depend on the photoexcitation density and the Al content
$x\in (0,1]$ in the SL barriers. For high photoexcitation densities,
$\delta\ll 1$, whereas $\delta\gg 1$ for small photoexcitation densities.
The parameter $\nu$ can be of any order.

It is interesting to depict the phase plane corresponding to spatially uniform
solutions of Eqs.~(\ref{eq1})--(\ref{eq3}) with $n_{i}=p_{i}=p$, $F_{i}=F$:
\begin{eqnarray}
\delta\,\frac{dF}{dt} =J-p\, v(F), \quad\mbox{and}\quad\frac{dp}{dt}=1 - r(F)\, p^2\;.
\label{eq5}
\end{eqnarray}
Generically and for fixed $J$, the nullclines $v(F)\, p=J$ and $r(F)\, p^2=1$
intersect in one or three fixed points, depending on the Al content $x$ in the barriers,
a parameter that controls the functions $v(F)$, $D(F)$ and $r(F)$.\cite{Ara10}
At these fixed points,
\begin{equation}
j(F) = J, \quad j(F)=\frac{v(F)}{\sqrt{r(F)}}\;.
 \label{eq6}
\end{equation}

\begin{figure}[!b]
\centerline{\includegraphics[width=0.5\linewidth]{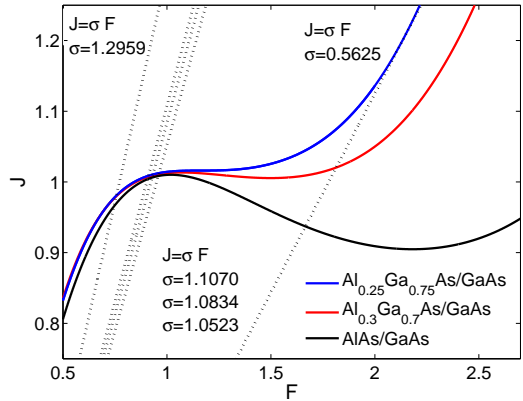}}
\caption{(Color online) Local current density $j$ vs electric field $F$. For a fixed value
of the total current density $J$, there may be one or three zeros of $j(F)-J$
depending on the Al content $x$. The dashed lines are the boundary condition $J=\sigma F$ for the different values of the dimensionless contact conductivity $\sigma$ indicated in the figure.}  
\label{fig2}
\end{figure}
The function $j(F)$ is depicted in Fig.~\ref{fig2}. For $0.45\leq x\leq 1$,
there are three fixed points of the system in Eq.~(\ref{eq5}), one on each of
the three branches of $p=J/v(F)$, and the ratio between
$j_{\rm max}-j_{\rm min}$ and the average current $(j_{\rm max}+j_{\rm min})/2$
is sufficiently large.
In this case, nonlinear phenomena are quite similar to those observed in
doped SL: static electric field domains with domain walls joining the
stable branches of $p=J/v(F)$, SSOC in voltage-biased SL due to the recycling
of pulses formed by two moving domain walls bounding a high field region,
etc. For $0<x<0.45$, $j(F)$ is either increasing for positive $F$ (if
$0<x< 0.25$) or the ratio
$2(j_{\rm max}-j_{\rm min})/(j_{\rm max}+ j_{\rm min})$ is small
(if $0.25<x<0.45$). For $0<x<0.25$, there is a unique fixed point at
$F=F_*$, which for an appropriate value of $J$ may be located at any of
the three branches of $p=J/v(F)$. If the fixed point is located at one of the
two stable branches of $p=J/v(F)$ for which $v(F)$ has a positive slope, the
dynamical system of Eq.~(\ref{eq5}) is excitable, whereas it is oscillatory if
the fixed point is located on the second branch of $v(F)$ with negative slope.
We shall see later that quite unusual phenomena are found for these cases.
Under dc current bias, it is possible to have pulses moving to the right
or to the left and periodic wave trains. Under dc voltage bias, these pulses
and wave trains may give rise to SSOC.\cite{Ara10}

\section{Wave fronts in a dc-current-biased photoexcited SL behaving as an
excitable medium}
\label{sec:wf}
Wave fronts and pulses of the electric field and carrier densities are key elements in the description of stable solutions of our model equations. We will start by describing these solutions for an infinite SL under constant current bias $J$ in the limit of high photoexcitation densities, $\delta\ll 1$, in which the dynamical behavior of the SL is richer.\cite{Ara10} In this Section, we focus our attention on wave fronts, while we will deal with pulses in the next Section.

\subsection{Leading order construction of wave fronts}
A wave front is a moving interface separating regions of smooth field variation
on the time scale $t$. Inside the front, $F_{i}$ varies rapidly on the time
scale $t/\delta$. Let us eliminate the electron density $n_i$ by using
Eq.~(\ref{eq1}) in Eqs.~(\ref{eq2})--(\ref{eq3}):
\begin{eqnarray}
\delta\, \frac{dF_{i}}{dt} &+& v(F_i)\,\frac{F_{i}-F_{i-1}}{\nu}- D(F_i)\,
\frac{F_{i+1}+F_{i-1}-2F_i}{\nu} \nonumber\\
&=& J- v(F_i)p_i + D(F_i)\, (p_{i+1}-p_i)\;,
\label{eq7}\\
\frac{dp_{i}}{dt}&=&1 - r(F_{i})\, p_{i}\left(p_{i}+
\frac{F_{i}-F_{i-1}}{\nu}\right)\;,
\label{eq8}
\end{eqnarray}
In the regions where the field varies smoothly, we can set $\delta=0$,
$F_{i}= F_{i-1}$, $n_i=p_i$ in Eqs.~(\ref{eq7})--(\ref{eq8}), thereby
obtaining the reduced problem
\begin{eqnarray}
&& p_{i} v(F_{i}) = J\;,
\label{eq9}\\
&&\frac{dp_{i}}{dt} = 1-r(F_{i})\, p_{i}^2\;.
\label{eq10}
\end{eqnarray}
In the wave fronts that separate these regions, the electric field and the hole
density vary rapidly as $F_{i}(t)=F(\xi)$, $p_{i}(t)=p(\xi)$, with $\xi=i-ct/\delta$.
In these regions, Eqs.~(\ref{eq7}) and (\ref{eq8}) yield to leading order
\begin{eqnarray}
 -c\,\frac{dF}{d\xi} &=& J-\left[p+\frac{F(\xi)-F(\xi-1)}{\nu}\right]v(F(\xi))
\nonumber\\
&+& D(F(\xi))\, \frac{F(\xi+1)+F(\xi-1)-2F(\xi)}{\nu}\;,
\label{eq11}\\
-c\,\frac{dp}{d\xi}&=& 0\;.
\label{eq12}
\end{eqnarray}
Thus $p$ is a constant equal to the value $p_i(t)$ at the last point in the region of
smooth variation before the front. Let $F^{(1)}(p)<F^{(2)}(p)<F^{(3)}(p)$ be the
solutions of $J/v(F)=p$ for $v_{\rm min}<J/p<v_{\rm max}$, where the local maximum
and minimum of the velocity $v(F)$ are reached at $(F_{\rm max}, v_{\rm max})$ and
at $(F_{\rm min},v_{\rm min})$, respectively (with $F_{\rm max}<F_{\rm min}$).
Let us assume that the front velocity is positive, $c>0$. Equation~(\ref{eq11})
has {\em decreasing} front solutions (DFs) such that the profile
$F(i-ct/\delta)$ is a decreasing function satisfying the boundary conditions:
\begin{eqnarray}
 F(-\infty)= F_{i}(+\infty)=F^{(3)}(p), \quad F(+\infty)= F_{i}(-\infty)=F^{(1)}(p)\;, \label{eq13}
\end{eqnarray}
and {\em increasing} front solutions (IFs) with an increasing profile $F(i-ct/\delta)$
that satisfies the boundary conditions:
\begin{eqnarray}
F(-\infty)= F_{i}(+\infty)= F^{(1)}(p), \quad F(+\infty)= F_{i}(-\infty)=F^{(3)}(p)\;. \label{eq14}
\end{eqnarray}

\begin{figure}[!b]
\centerline{\includegraphics[width=0.5\linewidth]{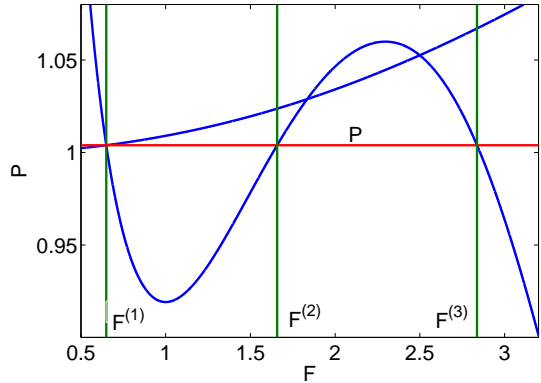}}
\caption{(Color online) $p$-$F$ phase plane exhibiting nullclines for fixed $J$ and a wave
front at constant $p$ in a GaAs/AlAs SL.}
\label{fig3}
\end{figure}
In Fig.~\ref{fig3}, we show the nullclines $p=J/v(F)$ and $p=[r(F)]^{-1/2}$
for fixed $J$ corresponding to Eq.~(\ref{eq5}) in a GaAs/AlAs SL. Depending on
the parameters, the increasing function $p=[r(F)]^{-1/2}$ may intersect the cubic
$p=J/v(F)$ in one, two or three critical points. At least one critical point
on the first or third branch of $p=J/v(F)$ makes our system behave as an
excitable medium, whereas if there is only one critical point on the second
branch of $p=J/v(F)$, our system has an oscillatory character, as we will
see later. In Fig.~\ref{fig3}, we have also shown a horizontal line for a
given constant value of $p$, which joins the first and third branches of
$p=J/v(F)$. This line corresponds to a wave front with fixed $p$ as
explained above.

\begin{figure}[!t]
\centerline{\includegraphics[width=0.5\linewidth]{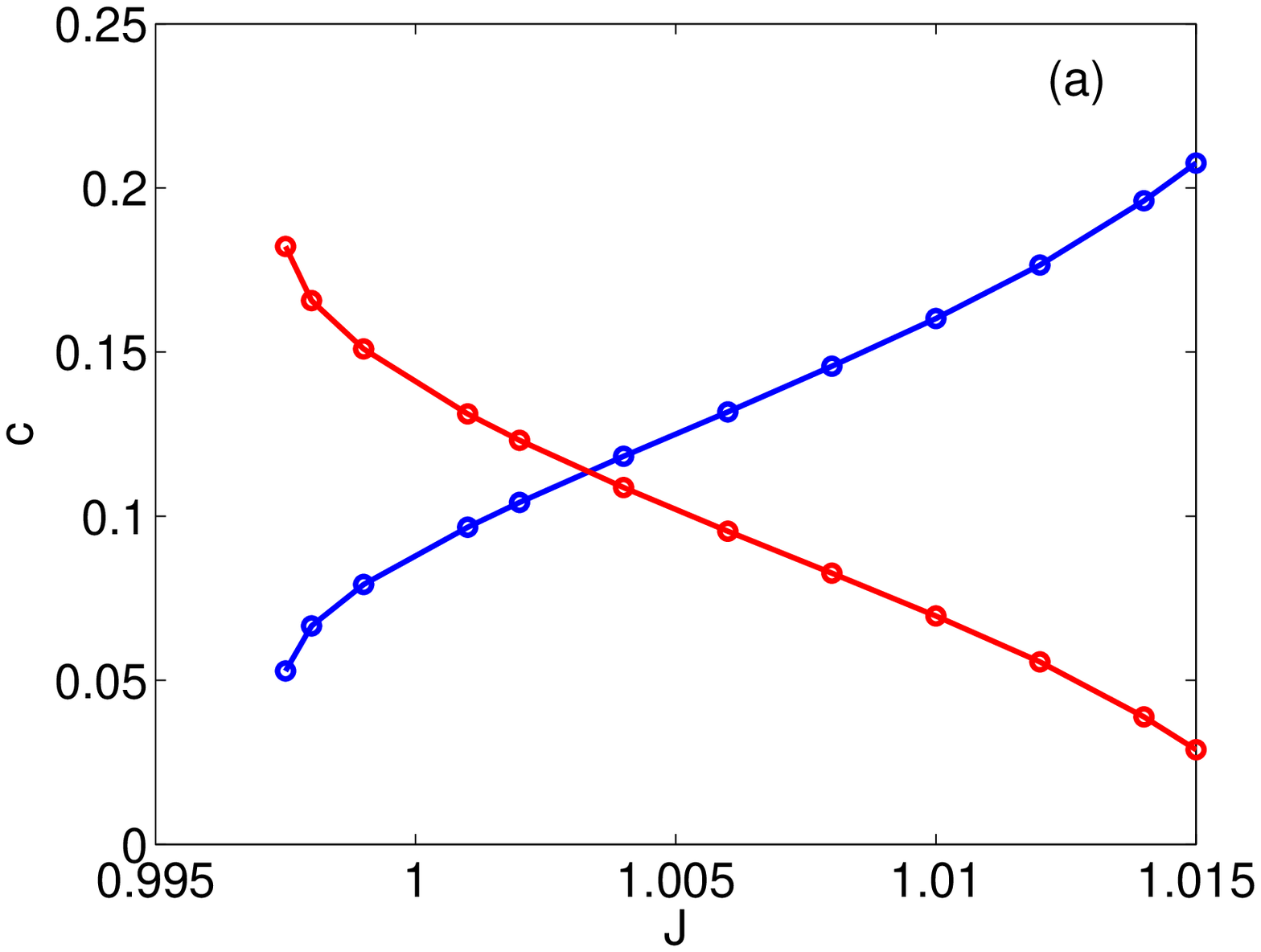}}
\centerline{\includegraphics[width=0.5\linewidth]{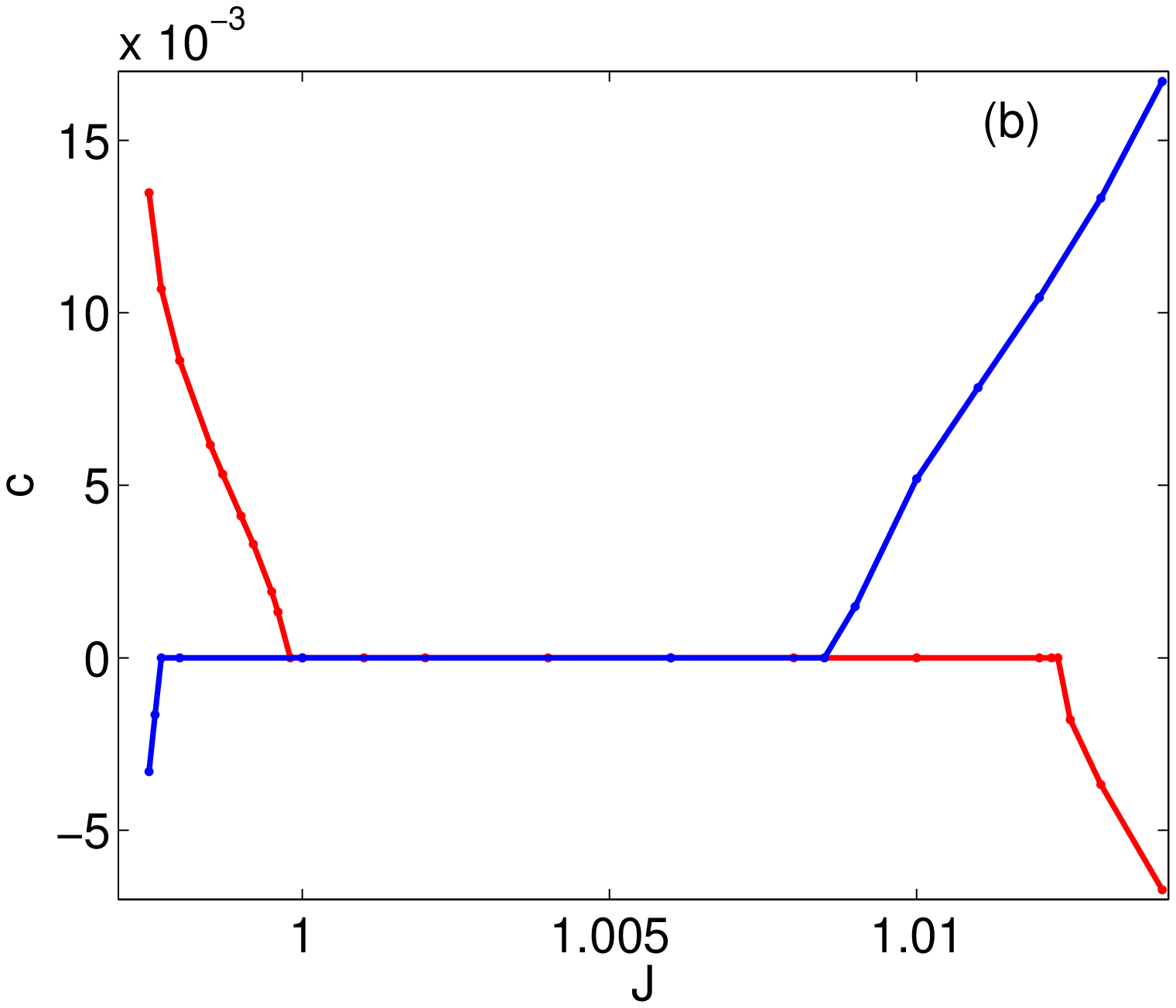}}
\caption{(Color online) Dimensionless velocities of DF and IF as a function of dimensionless current density $J$ for (a) $\nu=8.45$ and (b) $\nu=330$ with $x=0.25$ and $p$ fixed. The DF (resp. IF) velocity increases (resp. decreases) with increasing $J$.}
\label{fig4}
\end{figure}

For sufficiently large $\nu$, there are two critical values of the current density,
$J_{c1}$ and $J_{c2}$, so that IFs move to the right ($c>0$) if $J<J_{c1}$, are
pinned ($c=0$) if $J_{c1}<J<J_{c2}$, and move to the left ($c<0$) if $J>J_{c2}$,
as shown in Fig.~\ref{fig4}(a). A similar picture holds for DFs [cf. Fig. \ref{fig4}(b)].
Near the critical currents $J_{ci}$ ($i=1,2$), the wave front profiles and their
velocities can be approximately found by means of the theory of active quantum
wells (QWs) developed for doped SLs in Ref.~\cite{CBA} (the only change is
replacing $J/p$ and $\nu/p$ instead of $J$ and $\nu$ in the expressions for doped SLs).
In the limit $\nu\to 0+$, the equations for the fronts can be approximated by their
continuum limit, which is a first-order hyperbolic partial differential equation having shock waves among its solutions, and the shock velocity gives a good approximation of the wave front velocity.\cite{CBA}

The theory we have just sketched holds in the limit as $\delta\to 0$, and we would
expect that it also holds for sufficiently small positive values of $\delta$.
However, a comparison of the front velocity given by the asymptotic theory with the
one obtained by direct numerical simulation shows a remarkable difference even for
quite small values of $\delta$. The difference between the approximate and
numerical velocities for $\delta=0.018$ is about 0.06, which gives a relative
error of 25\%. This error is reduced to 2\% for $\delta= 0.001$. Nevertheless,
we would like to have an asymptotic theory, which is better than this. How do
we correct the simple wave front construction given above?

\subsection{Corrected asymptotic theory of wave fronts}
To correct the previous leading order theory of wave fronts, we use the fact
that the DF or IF profiles are monotone functions of a variable $\xi=i-ct/\delta$. Thus
\begin{eqnarray}
F_{i}(t)= \mathcal{F}(\xi),\quad p_{i}(t)=\mathcal{P}(\xi)\;,
\label{eq15}
\end{eqnarray}
with $\mathcal{F}'(\xi)<0$ (resp. $>0$) for DF (resp. IF). In either case, we can find
\begin{eqnarray}
\xi= \Xi(F)\quad\mbox{which solves }\quad \mathcal{F}(\xi)=F\;.
\label{eq16}
\end{eqnarray}
Then the finite differences
\begin{eqnarray}
&& F_{i}-F_{i-1}= \mathcal{F}(\xi)-\mathcal{F}(\xi-1)=F_{i}-\mathcal{F}(\Xi(
F_{i})-1),\quad\mbox{and} \nonumber\\
&& F_{i+1}+F_{i-1}-2F_{i}= \mathcal{F}(\Xi(F_{i})+1)+\mathcal{F}(\Xi(F_{i})-1)-
2F_{i}
\label{eq17}
\end{eqnarray}
can be considered to be functions of $F_{i}$. Therefore, we can derive from the
equation for $p_{i}(\xi)$
\begin{eqnarray}
-c\frac{dp_{i}}{d\xi} =\delta\left[ 1-r(F_{i})\, p_{i}\left(p_{i}+\frac{F_{i}-
F_{i-1}}{\nu}\right)\right]
\label{eq18}
\end{eqnarray}
and from the equation for the wave front field profile the following equation
for $p_{i}$ as a function of the field $F_{i}$:
\begin{eqnarray}
\frac{dp_{i}}{dF_{i}} &=& \frac{\delta}{J}\frac{1-r(F_{i})\, p_{i}\left(p_{i}+
\frac{F_{i}-F_{i-1}}{\nu}\right)}{1-\left(p_{i}+\frac{F_{i}-F_{i-1}}{\nu}
\right)\frac{v(F_{i})}{J} + \frac{D(F_{i})}{J}\left(p_{i+1}-p_{i}+
 \frac{F_{i+1}+F_{i-1}-2F_{i}}{\nu}\right)}\;.
\label{eq19}
\end{eqnarray}
Now we integrate this equation and iterate the result starting from the
value $p_{i}(F_{0})=p$, thereby obtaining
\begin{eqnarray}
&& p_{i}\sim p \nonumber\\
&& + \frac{\delta}{J}\int_{F_{0}}^{F_{i}}\frac{\left[1-r(F)p
\left(p+\frac{F-\mathcal{F}(\Xi(F)-1)}{\nu}\right)\right]\, dF}{1-\left(p+
\frac{F-\mathcal{F}(\Xi(F)-1)}{\nu}\right)\frac{v(F)}{J} + \frac{D(F)}{J}
\left( \frac{\mathcal{F}(\Xi(F)+1)+\mathcal{F}(\Xi(F)-1)-2F}{\nu}\right)}
\label{eq20}
\end{eqnarray}
up to terms of order $\delta^2$.

The starting point $F_{0}$ should be selected so as to ensure convergence
of the integral in Eq.~(\ref{eq20}). If $p=p_{*}$ corresponds exactly to
a critical point, we can select $F_{0}$ as the field value of the same
critical point $F_{*}$. Then the integrand has a finite limit as $F\to F_{0}$,
and the integral in Eq.~(\ref{eq20}) converges. If this is not the case,
we need a nonzero value of the finite difference
$F_{0}-\mathcal{F}(\Xi(F_{0})-1)$ to obtain a nonzero value of the denominator
in Eq.~(\ref{eq19}). We use the slow scale equation (\ref{eq9}) to
calculate $F_{i}=\Phi(p_{i})$. Inserting this function in Eq.~(\ref{eq10}),
we obtain
\begin{eqnarray}
\frac{dp_{i}}{dt} = 1-r(\Phi(p_{i}))p_{i}^2\;.    \label{eq21}
\end{eqnarray}
We now solve this equation for an initial value $p_{i}(0)=p$ to obtain
$p_{i}=p_{i}(t;p)$. Then $F_{i}(t)=\Phi(p_{i}(t;p))$. Using that $F_{i}$
is a function of $i-ct/\delta$ in the wave front profile,
we have $F_{i-1}(t=0)=F_{i}(t=-\delta/c)=\Phi(p_{i}(-\delta/c;p))$. The equation
\begin{eqnarray}
\mathcal{F}(\Xi(F_{0})-1) =\Phi\left(p_{i}\left(-\frac{\delta}{c};p\right)
\right) \label{eq22}
\end{eqnarray}
determines $F_{0}$ as a function of $p$ with $F_{0}-\mathcal{F}(\Xi(F_0)-1)\neq 0$.

\begin{figure}[!t]
\centerline{\includegraphics[width=0.5\linewidth]{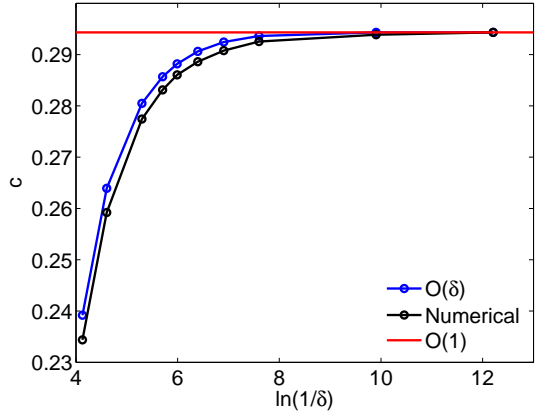}}
\caption{(Color online) Comparison of the numerically obtained and asymptotic approximations for the DF velocity as a function of $\ln(1/\delta)$ using the parameter values $x=0.3$, $J=1.006$, and $\nu= 4.449$.}
\label{fig5}
\end{figure}

Once we have determined $p_{i}$ as a function of $F_{i}$ by means of Eq.~(\ref{eq20}),
we can solve the fast equation
\begin{eqnarray}
-c\,\frac{dF_{i}}{d\xi} &+&\left(p_{i}(F_{i})+\frac{F_{i}-F_{i-1}}{\nu}\right)
v(F_{i}) \nonumber\\
&-& D(F_{i})\left(p_{i+1}(F_{i+1})-p_{i}(F_{i})+\frac{F_{i+1}+F_{i-1}-2F_{i}}{
\nu}\right) = J \label{eq23}
\end{eqnarray}
with $F_{i}=F(\xi)$, $F_{i\pm 1}=F(\xi\pm 1)$, and boundary conditions (for $c>0$)
\begin{eqnarray}
F_{i}(-\infty)= F(+\infty)= F^{(3)}(p'), \quad\mbox{and}\quad
F_{i}(+\infty)= F(-\infty)= F^{(1)}(p), \label{eq24}
\end{eqnarray}
for the IFs and
\begin{eqnarray}
F_{i}(-\infty)= F(+\infty)= F^{(1)}(p'), \quad\mbox{and}\quad
F_{i}(+\infty)= F(-\infty)= F^{(3)}(p), \label{eq25}
\end{eqnarray}
for the DFs. In Eq.~(\ref{eq24}), the value $p'\neq p$ is determined by solving
Eqs.~(\ref{eq20}) and (\ref{eq23}), until the first branch of $J/v(F)$ is reached.
Similarly, for a DF, $p'$ in Eq.~(\ref{eq25}) is found by solving Eqs.~(\ref{eq20})
and (\ref{eq23}), until the third branch of $J/v(F)$ is reached. The front velocity
is now a function of $\delta$, and Fig.~\ref{fig5} shows that it is a much better
approximation to the numerically calculated front velocity than that given by the
leading order theory. In practice, it is easier to calculate directly the finite
differences $\mp(F_{i}-F_{i\pm 1})$ as functions of $F_{i}$.

\section{Pulses moving downstream in an excitable SL}
\label{sec:pulses}
\begin{figure}[!t]
\centerline{\includegraphics[width=0.5\linewidth]{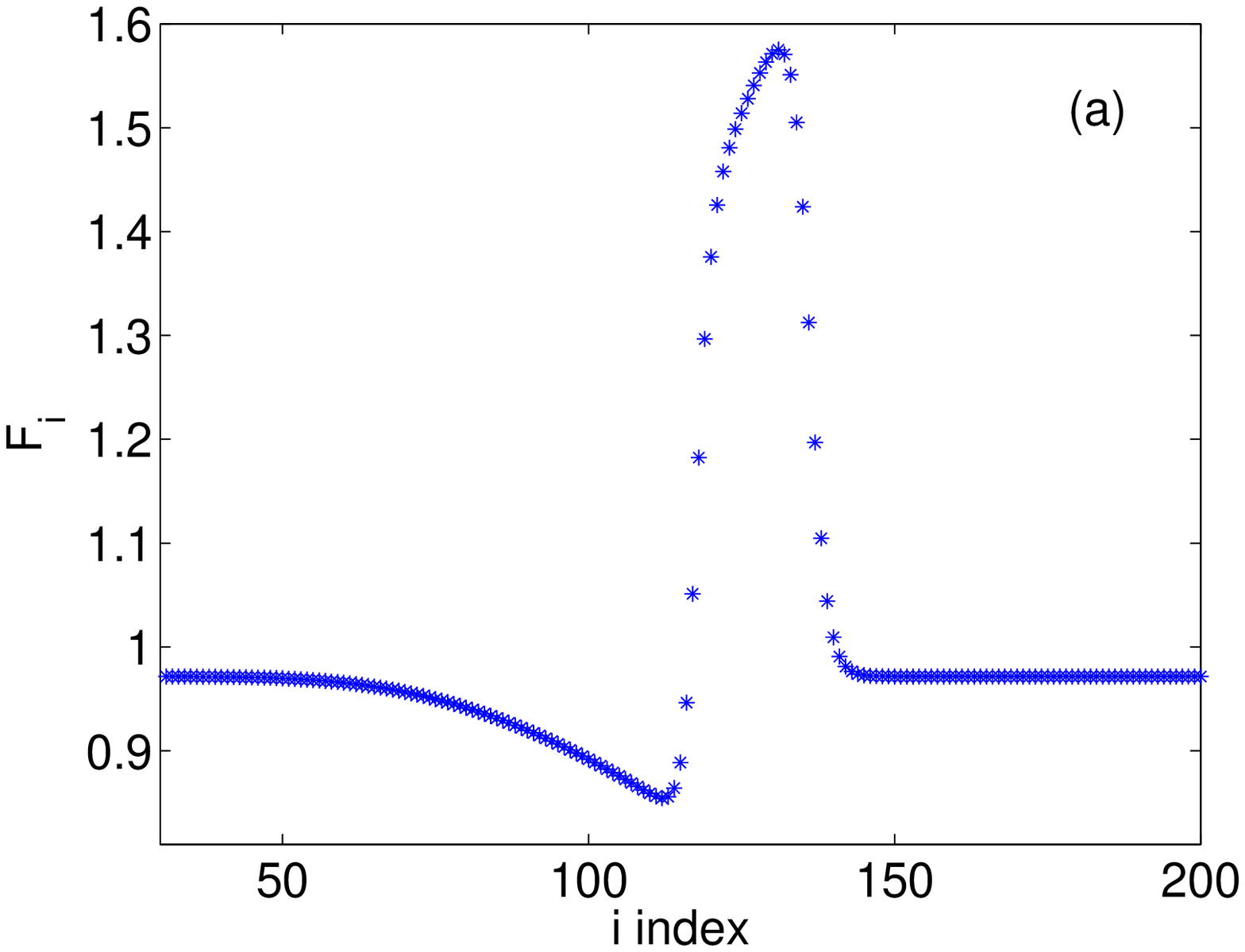}}
\centerline{\includegraphics[width=0.5\linewidth]{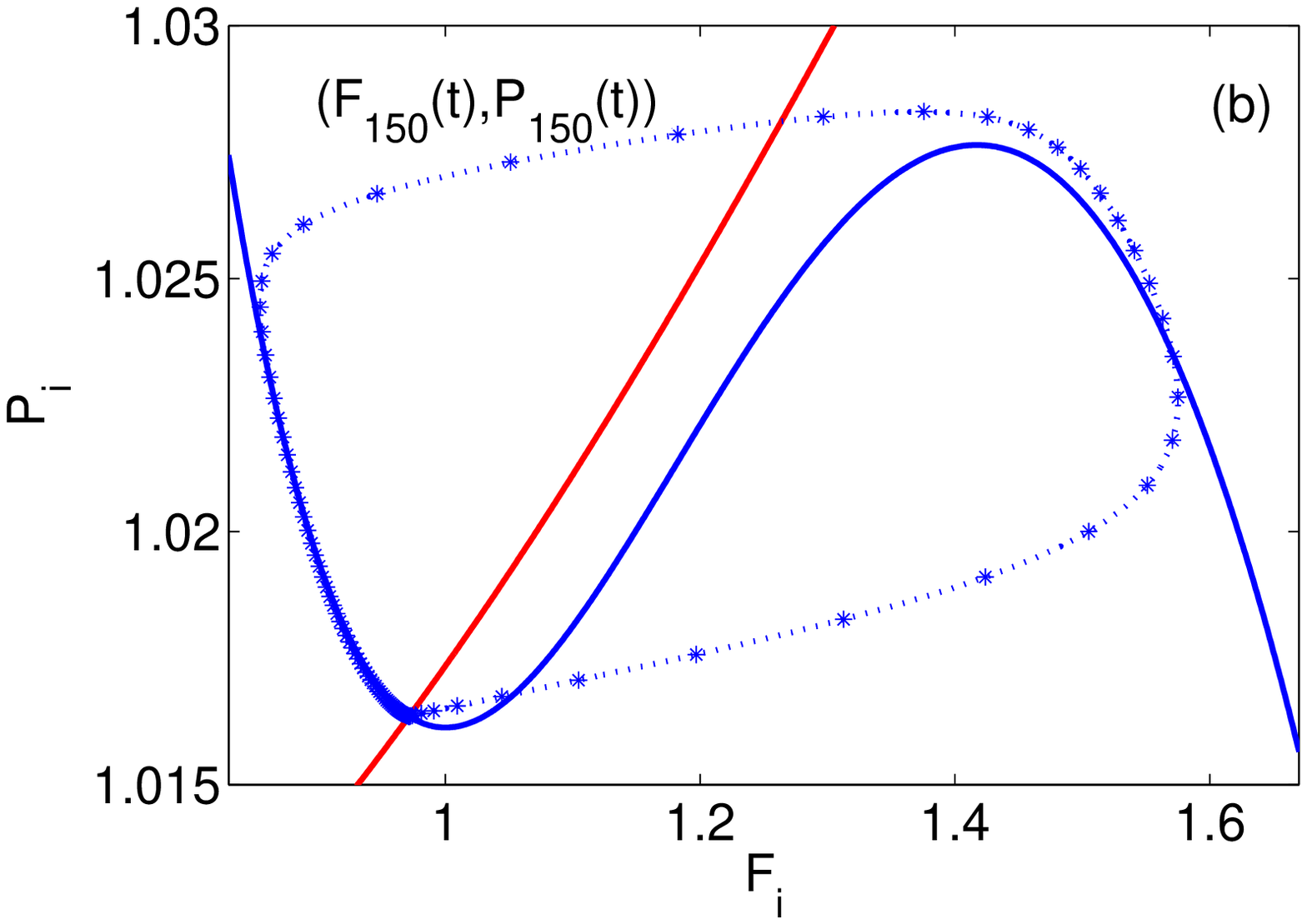}}
\caption{(Color online) (a) Numerically obtained field profile of a pulse moving with
positive velocity for $J=1.013$, $\nu=8.44565$, $\delta=3.7288\times 10^{-3}$.
(b) Phase plane showing the nullclines and the motion of the 150th QW as the
pulse traverses it. The laser intensity is 120.5 kW/cm$^2$. }
\label{fig6}
\end{figure}

As we see in Fig.~\ref{fig6}(a), a pulse moving downstream with {\em positive}
velocity consists of regions of smooth field variation on the time scale $t$,
separated by sharp interfaces in which $F_{i}$ varies rapidly on the time scale
$t/\delta$. To find an asymptotic approximation to the pulse profile, we first
use the leading order description of its component wave fronts, according to
which $p_{i}$ is a constant independent of $i$ inside the wave front. We can
now discuss different regions in the asymptotic description of a pulse,
recalling that the field profile is the mirror image of the motion of a
QW, $F_{i}(t)=F(i-ct/\delta)$.

\subsection{Only one critical point on a stable branch of $J/v(F)$
(leading order theory)}
First of all, we shall describe the pulses for values of $J$ and $\nu$ such
that there is one critical point $(F_{*},p_{*})$ located on the first branch
of $J/v(F)$. There are no other critical points or, if they exist, they are
located on the second branch of $J/v(F)$. Similar results are obtained if
the critical point is on the third branch of $J/v(F)$.

\begin{itemize}
\item The region of smooth variation of $F$ in front of the pulse is described
by Eqs.~(\ref{eq9}) and (\ref{eq10}). In this region, $F_i = F^{(1)}(p_i)$ so
that
$$\frac{dp_{i}}{ dt} = 1 - r\left(F^{(1)}(p_i)\right) p_{i}^2\;,$$
and initial data evolve exponentially fast toward equilibrium,
$F_i = F_{*},\, p_i = p_{*}$.

\item For the pulse leading edge, let $p(t)$ be the value of $p_i$ at
the last point of the region in front of the pulse. Eventually, $p\to p_{*}$.
At the leading edge, $F_i(t) = F(i-ct/\delta)$ is a DF moving toward the right
with velocity $C= c(J,\nu,p)/\delta$ measured in QWs per unit time $t$. The DF
satisfies Eq.~(\ref{eq11}) with the boundary conditions in Eq.~(\ref{eq13}).
It is convenient to call $c_-(p)= c(J,\nu,p)$. Eventually,
$C\sim c_-(p_{*})/\delta$, and $F_i$ decreases from $F_i=F_{\rm max}$ to
$F_i =F_{*}$ across the leading edge of the pulse.

\item For the region between fronts, $F_i=F^{(3)}(p_i)$ and
$$\frac{dp_{i}}{ dt} = 1 - r\left(F^{(3)}(p_i)\right)\, p_{i}^2\;.$$
There is a finite number of points in this region. On its far right,
$p_i=p\to p_{*}$. As we move toward the left, $p_i$ increases until it reaches
a certain value $P(t)$ corresponding to that in the trailing wave front.

\item For the trailing wave front, $p_i(t)=p(\xi)=P$, and $F_i(t)=F(\xi)$ is
an IF satisfying Eq.~(\ref{eq11}) with the boundary conditions of
Eq.~(\ref{eq14}). This front moves with velocity $C= c(J,\nu,P)/\delta$
measured in QWs per unit time $t$. It is convenient to denote
$c_+(P)=c(J,\nu,P)$. We shall indicate how to determine $P$ below. Clearly,
if the pulse moves rigidly, we should have $c_+(P)=c_-(p_{*})$ after a
sufficiently long transient period.

\item For the pulse tail, we again have $F_i=F^{(1)}(p_i)$ and
$dp_{i}/dt=1- r(F^{(1)}(p_{i}))\,p^2_{i}$. Sufficiently far to the
left, $p_i=p_{*},\, F_i=F_{*}$.
\end{itemize}

The number of QWs between wave fronts of the pulse can be calculated as
follows.\cite{Car03} Let $\tau$ be the delay between fronts, i.e. the time
elapsed from the moment, at which the leading front traverses the QW $i=I$,
to the moment, when the trailing front is at $i=I$. Clearly,
\begin{eqnarray}
\tau =\int_{p(t-\tau)}^{P(t)} \frac{dp}{1-  r\left(F^{(3)}(p)\right)\, p^2}\;.
\label{eq26}
\end{eqnarray}
The number of QWs between fronts, $m(t)$, is
\begin{eqnarray}
m =\frac{1}{\delta}\,\int_{t-\tau}^{t} c_-(p(t))\, dt\;.   \label{eq27}
\end{eqnarray}
However, the separation between fronts satisfies the equation
\begin{eqnarray}
\frac{dm}{ dt} = \frac{c_{-}(p(t)) - c_{+}(P(t))}{ \delta};.   \label{eq28}
\end{eqnarray}
The three Eqs.~(\ref{eq26}), (\ref{eq27}), and (\ref{eq28}) can be solved to
obtain the three unknowns $\tau$, $m$ and $P(t)$. The function $p(t)$ is
determined by solving Eq.~(\ref{eq10}) with $F_i= F^{(1)}(p_i)$ in the region to
the left of the leading front.

After a transient period, $p(t)\to p_{*}$ and $P(t)\to P$ (a constant value)
so that we arrive at the simpler expressions
\begin{eqnarray}
&&\tau =\int_{p_{*}}^{P} \frac{dp}{1-  r\left(F^{(3)}(p)\right)\, p^2},\quad\mbox{and} \label{eq29}\\
&& \frac{dm}{ dt} = \frac{c_{-}(p_{*}) - c_{+}(P)}{ \delta} \label{eq30}
\end{eqnarray}
instead of Eqs.~(\ref{eq26}) and (\ref{eq28}), respectively. The number of
points at the pulse top is now
\begin{eqnarray}
m= \frac{c_-(p_{*})\tau}{\delta} = \frac{c_-(p_{*})}{\delta}\,
\int_{p_{*}}^{P} \frac{dp}{1-  r(F^{(3)}(p))\, p^2}\;. \label{eq31}
\end{eqnarray}
This equation yields $P$ as a function of $m$. Then Eq.~(\ref{eq30}) becomes
an autonomous differential equation for $m$ that has a stable constant
solution at $m=m^*$ such that $c_-(p_{*})= c_+(P(m))$: At $m=m^*$, the right
hand side of Eq.~(\ref{eq30}) has a slope
$-[1-  r(F^{(3)}(J/P))\, P^2]\, c'_+(P)/ c_-(p_{*}) <0$.

Recapitulating, for appropriate initial conditions, leading and trailing fronts
of a pulse evolve until $m$ reaches its stable value at which
$c_-(p_{*})= c_+(P(m^*))$ and Eq.~(\ref{eq31}) holds. To compute $m^*$, we first
determine $P^*=P(m^*)$ by using $c_-(p_{*}) =c_+(P(m^*))$. Then we calculate
$\tau=\tau^*$ (which does not depend on $\delta$!) from Eq.~(\ref{eq29}) and
$m^*=c_-(p_{*})\tau^*/\delta$. Our construction breaks down if the number of
QWs between fronts falls below 1. This yields an upper bound for the critical
value of $\delta$ above which pulse propagation fails: $\delta_c \sim c_-(p_{*})\tau^*$.
Pulse propagation may also fail if the leading front becomes pinned for a current
density on the interval $(J_{c1},J_{c2})$ as mentioned in the previous Section.\cite{Bon10,CBA,CBprl03,Wan08}

To calculate the asymptotic length of the pulse tail, we cannot use Eq.~(\ref{eq29})
with $F^{(1)}$ replacing $F^{(3)}$ in that formula, because the resulting time
becomes infinite. However, we can calculate the time it takes for a solution
$p(t)$ of Eq.~(\ref{eq21}) with $\Phi(p)=F^{(1)}(p)$ and $p(0)=P$ to reach a
neighborhood of $p_*$. From that equation, we obtain
$$t=\int_p^P\frac{dp}{r(F^{(1)}(p))p^2-1}\sim \int_{p_*}^P
\left[\frac{1}{r(F^{(1)}(p))p^2-1}-\frac{1}{(rp^2)'_*(p-p_*)}\right]dp+
\frac{1}{(rp^2)'_*}\ln\!\left(\frac{P-p_*}{p-p_*}\right)\;,$$
with $(rp^2)'_*=d[r(F^{(1)}(p))p^2]/dp|_{p=p_*}>0$, and therefore
\begin{eqnarray}
p(t)-p_*\sim (P-p_*)e^{-(rp^2)'_*(t-T)},\quad T=\int_{p_*}^P\left[
\frac{1}{r(F^{(1)}(p))p^2-1}-\frac{1}{(rp^2)'_*(p-p_*)}\right]dp\;, \label{eq32}
\end{eqnarray}
for long times such that $p(t)$ is sufficiently close to $p_*$. The time needed
for $p(t)$ to go from $P$ to $p_*+(P-p_*)/e$ is then $T_*=T+1/(rp^2)'_*$.
The tail length is approximately given by $M= c_-(p_{*})T_*/\delta$.

\subsection{Only one critical point on a stable branch of $J/v(F)$ (corrected theory)}
How do we correct this pulse construction using our improved theory of wave fronts?
\begin{itemize}
\item The region of smooth variation in front of the pulse is as described above.
\item However, we have to use Eq.~(\ref{eq23}) instead of Eq.~(\ref{eq11})
including the boundary conditions of Eq.~(\ref{eq24}) or (\ref{eq25}) to construct
the leading and trailing wave fronts of the pulse. Let us assume that the pulse
moves from left to right. The leading wave front is a DF moving with velocity
$c_{-}(p)/\delta$ with $c_{-}(p)=c(J,\nu,p,\delta)$ given by Eq.~(\ref{eq23})
and the boundary conditions of Eq.~(\ref{eq25}). While $p(t)$ is the value at
the initial field on the first branch of $J/v(F)$, $p'(t)$ is the hole density
at the final QW of the DF which is on the third branch of $J/v(F)$. The time it
takes for a QW to move from $(F^{(1)}(p),p)$ to $(F^{(3)}(p'),p')$ is of order
$\delta$, and it needs to be considered when constructing the pulse.
\item In the region between leading and trailing fronts, $F_{i}=F^{(3)}(p_{i})$.
On its far right, $p_{i} =p'\to p'_{*}$, where we call $p'_{*}$ the value of
$p'$ corresponding to $p=p_{*}$. As we move toward the left, $p_{i}$ increases
until it reaches a certain value $P(t)$ corresponding to that in the trailing
wave front.
\item The trailing wave front is an IF moving with velocity $c_{+}(P)/\delta$,
with $c_{+}(P)=c(J,\nu,P,\delta)$ given by Eq.~(\ref{eq23}) and the boundary
conditions of Eq.~(\ref{eq24}).
The hole densities at the initial and final QWs of the IF are $P$ and $P'$,
respectively. The corresponding fields are $F^{(3)}(P)$ and $F^{(1)}(P')$,
respectively. Again the time it takes for a QW to move from $(F^{(3)}(P),P)$
to $(F^{(1)}(P'),P')$ is of order $\delta$ and will be included in our
calculations to order $\delta$.
\item The pulse tail is as described above except that its first QW has a
hole density $P'$ instead of $P$.
\end{itemize}
Equations (\ref{eq26}) to (\ref{eq28}) become
\begin{eqnarray}
&&\tau =\int_{p'(t-\tau)}^{P(t)} \frac{dp}{1-  r\left(F^{(3)}(p)\right)\, p^2}\;,
\label{eq33}\\
&& m =\frac{1}{\delta}\,\int_{t-\tau}^{t} c_-(p(t))\, dt\;,   \label{eq34}\\
&& \frac{dm}{ dt} = \frac{c_{-}(p'(t)) - c_{+}(P(t))}{\delta}\;.  \label{eq35}
\end{eqnarray}
After the transient period, these equations become Eqs.~(\ref{eq29})--(\ref{eq31})
with $p'$ (now time-independent) instead of $p_{*}$. The rest of the
considerations made above apply except that now $\tau^*$ depends on
$\delta$ because $p'_{*}$ does.

\begin{figure}[!t]
\centerline{\includegraphics[width=0.5\linewidth]{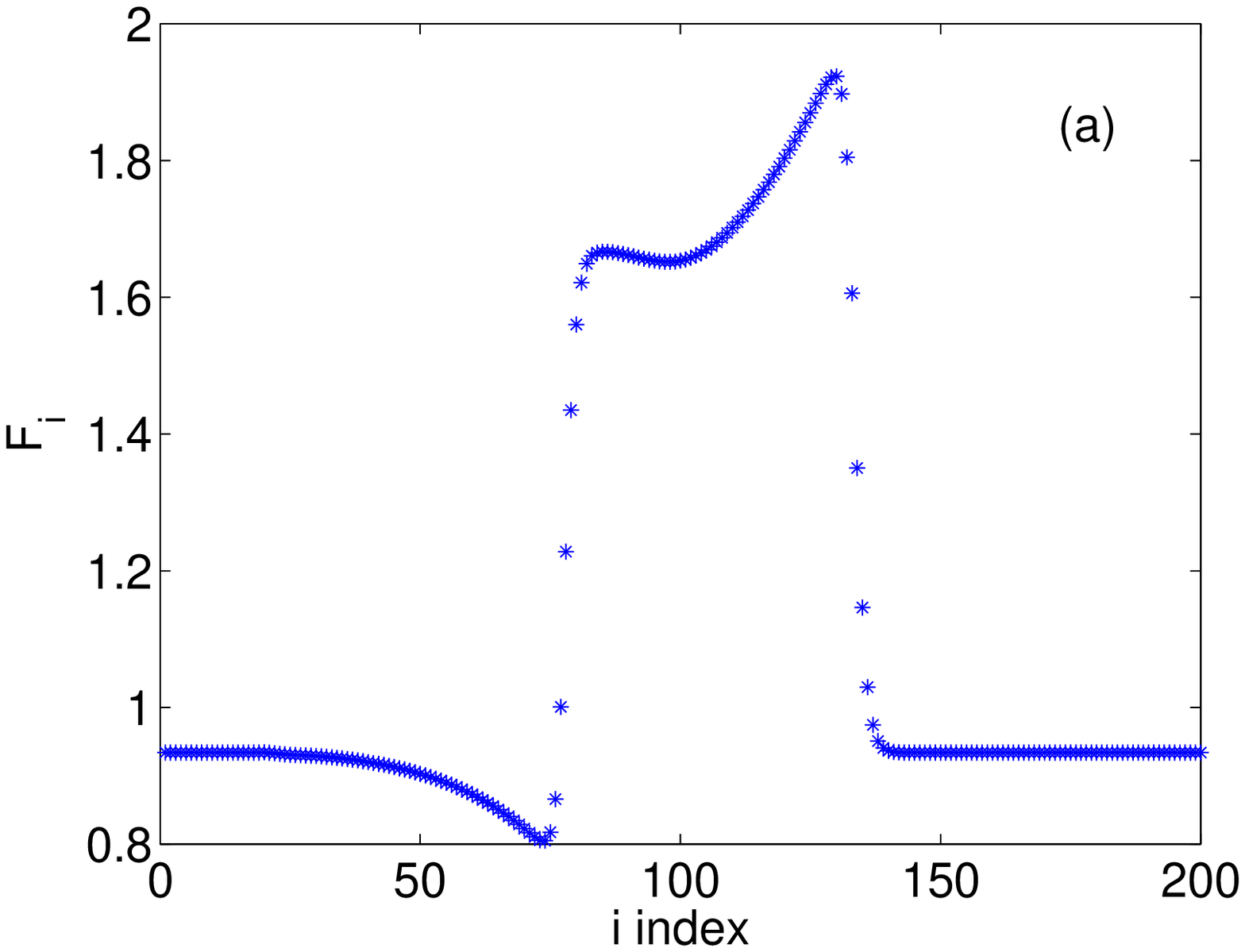}}
\centerline{\includegraphics[width=0.5\linewidth]{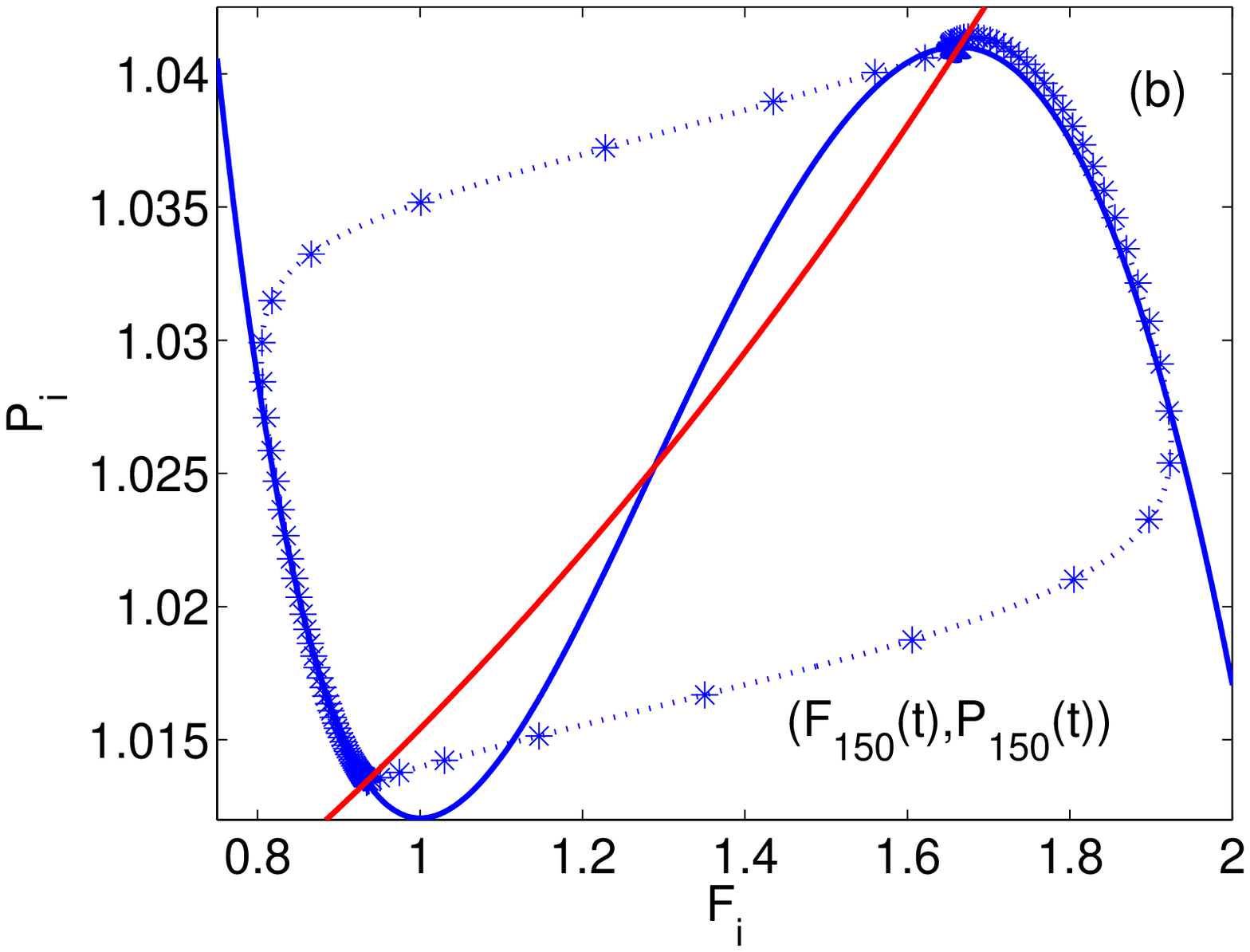}}
\caption{(Color online) (a) Numerically obtained field profile of a pulse moving with positive velocity when there are three critical points in the phase plane. (b) Phase
plane showing the nullclines and the motion of the 150$^{th}$ QW as the pulse
traverses it. $J=1.009$. The photoexcitation intensity is 120.5 kW/cm$^2$.}
\label{fig7}
\end{figure}

\subsection{Two critical points on the stable branches of $J/v(F)$}
In the case illustrated in Fig.~\ref{fig7}, there is one critical point on
each of the three branches of $J/v(F)$. Let $(F_{*},p_{*})$ and $(F^*,p^*)$
be the critical points on the first and third branches of $J/v(F)$, respectively.
This case is very similar to that described before for the case of a sole critical
point on the first branch. However, now $c_{-}(p_{*})\neq c_{+}(p^*)$ except for
particular values of $J$. Therefore, these pulses do not move rigidly in general:
they will shrink and disappear or grow indefinitely. Regions 1 to 3 of the pulse
in Fig.~\ref{fig7}(a) are identical to those of the pulse with only one critical
point, except that now $p\to p_{*}$ and $P\to p^*$, and $m$ either decays to zero
or it grows indefinitely.

\section{Wave trains moving downstream in a dc-current-biased photoexcited SL}
\label{sec:trains}
\begin{figure}[!t]
\centerline{\includegraphics[width=0.5\linewidth]{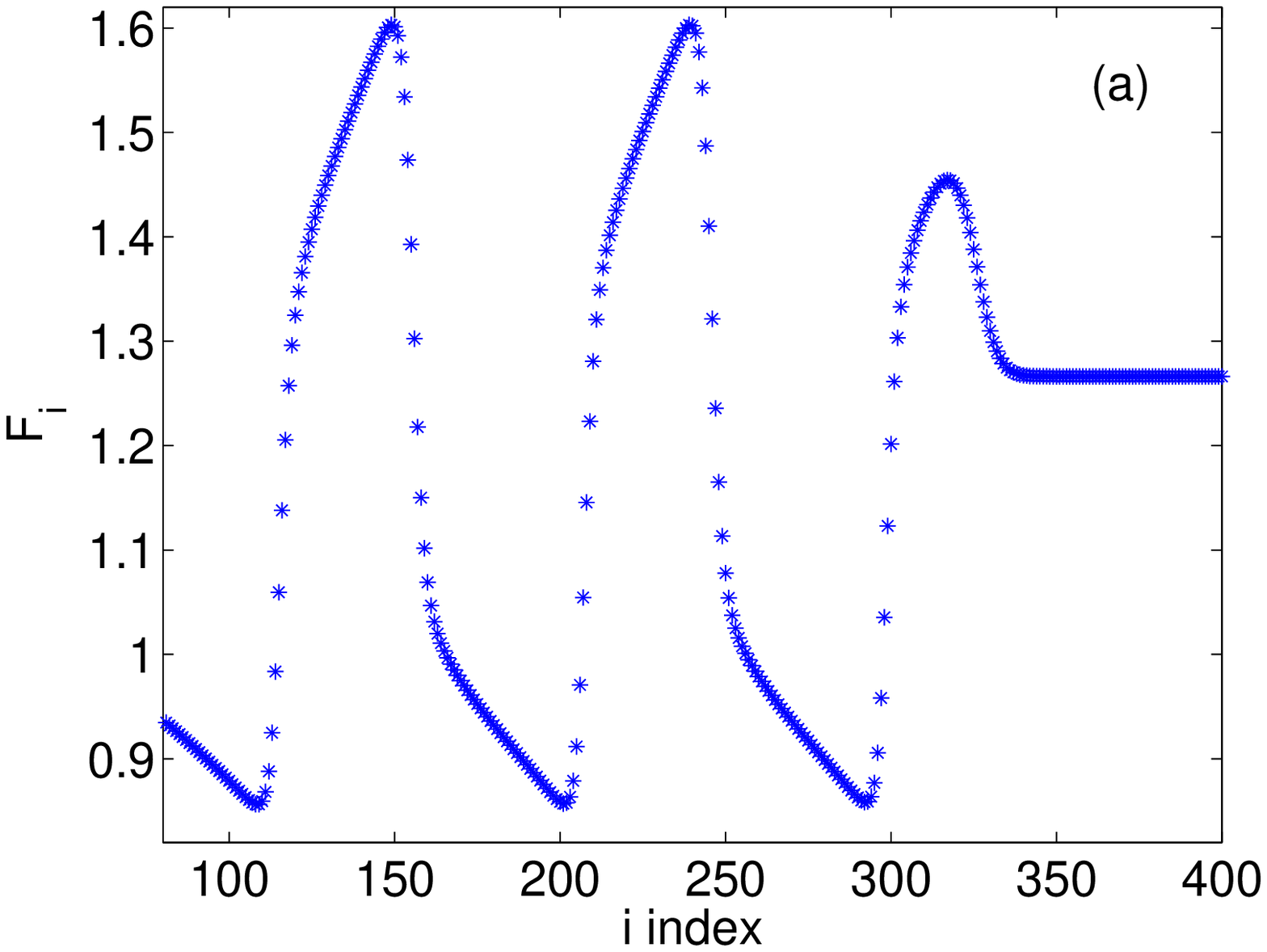}}
\centerline{\includegraphics[width=0.5\linewidth]{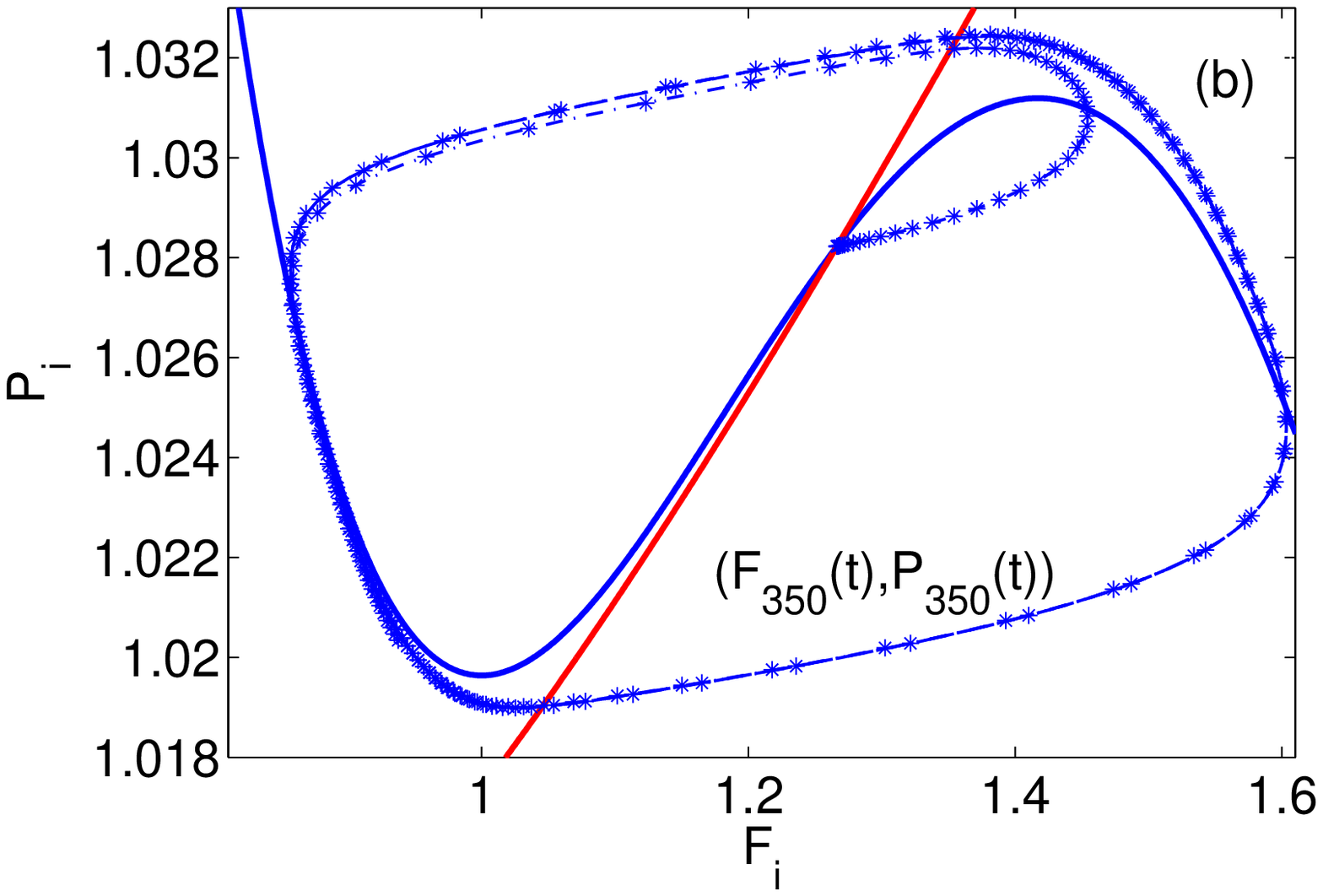}}
\caption{(Color online) (a) Numerically obtained field profile of a pulse moving with
positive velocity followed by a wave train when there is only one critical point
in the phase plane. (b) Phase plane showing the nullclines and the motion of
the 350$^{th}$ QW as the pulse traverses it for $J=1.0165$. The photoexcitation
intensity corresponds to 120.5 kW/cm$^2$.}
\label{fig8}
\end{figure}

For the spatially discrete FHN system, wave trains were constructed by
A. Carpio using matched asymptotic expansions.\cite{Car05} A similar construction
could be carried out using our improved theory of wave fronts. A wave train
consists of a periodic profile $F(\xi)$, $p(\xi)$, $\xi=i-ct/\delta$, with
period $L$ and velocity $c$. Figure~\ref{fig8}(a) gives the corresponding field
profile: a first smaller pulse triggers a periodic succession of equal pulses,
which become the wave train. Figure~\ref{fig8}(b) shows the passage of a wave
train through a QW: starting at an unstable stationary state, trajectories in
the phase plane evolve toward a stable limit cycle. The wave train profiles can
be reconstructed from a time periodic solution
$F_{i}(t)=F(i-ct/\delta)$, $p_{i}(t)=p(i-ct/\delta)$, with time period
$T=L\delta/c$ and velocity $c$. The spatial structure of the wave train at
each fixed time $t$ is given by $F_{i}(t)=F(i-ct/\delta)$ and
$p_{i}(t)=p(i-ct/\delta)$. The points contained in a period satisfy
$0 \leq i - ct/\delta\leq L$, that is, $ct/\delta\leq i\leq L + ct/\delta$.
As time grows, the discrete points \textit{travel} along the continuous wave
profile and are transferred from one period to the next. The number of integers
$i$ one can fit in an interval of length $L$ is the integer part of $L$.

Assuming a positive velocity, there are four stages in one time period of
the wave trains $(F_{i}(T,\delta),p_{i}(T,\delta))$:
\begin{itemize}
\item For leading-edge DFs given by Eqs.~(\ref{eq20}) and (\ref{eq23}) using
the boundary conditions of Eq.~(\ref{eq25}), the DF velocity is $c_{-}(p)/\delta$
with $c_{-}(p)=c(J,\nu,p,\delta)$, where $p(t)$ is the value at the initial
field on the first branch of $J/v(F)$ and $p'(t)$ is the hole density at the
final QW of the DF, whose corresponding field is on the third branch of
$J/v(F)$. The time it takes for a QW to move from $(F^{(1)}(p),p)$ to
$(F^{(3)}(p'),p')$ is of order $\delta$, and we should consider it when
constructing the train.
\item In the peak region between leading and trailing fronts, $F_{i}=F^{(3)}(p_{i})$.
On its far right, $p_{i} =p'$. As we move toward the left, $p_{i}$ increases
until it reaches a certain value $P(t)$ corresponding to that in the trailing
wave front. $P(t)$ will be calculated later. The duration of this stage is
\begin{eqnarray}
 T_{p} =\int_{p'}^{P} \frac{dp}{1-  r\left(F^{(3)}(p)\right)\, p^2}\;.
\label{eq36}
\end{eqnarray}
Note that $T_{p}$ may become infinite if there is a fixed point $(F^{*},p^{*})$
on the third branch of $J/v(F)$. In such a case, we may have wave trains only
if $p<p^{*}$.
\item The trailing wave front is an IF moving with velocity $c_{+}(P)/\delta$,
with $c_{+}(P)=c(J,\nu,P,\delta)$ given by Eq.~(\ref{eq22}) and the boundary
conditions of Eq.~(\ref{eq24}). The hole densities at the initial and final
QWs of the IF are $P$ and $P'$, respectively, and the corresponding fields
are $F^{(3)}(P)$ and $F^{(1)}(P')$, respectively. Again the time it takes for
a QW to move from $(F^{(3)}(P),P)$ to $(F^{(1)}(P'),P')$ is of order $\delta$
and will be ignored. For the wave train to move rigidly, we must have
$c_{-}(p)=c_{+}(P)= c$. This condition gives $P$ as a function of $p$.
\item In the tail of the wave train, $F_{i}=F^{(1)}(p_{i})$. On the far right
of the tail region, $p_{i}=P'$ and $p_{i}$ decreases until it reaches the
value $p$. The duration of this stage is
\begin{eqnarray}
 T_{t} =\int_{P'}^{p} \frac{dp}{1-  r\left(F^{(1)}(p)\right)\, p^2}\;.
\label{eq37}
\end{eqnarray}
Note that $T_{t}$ may become infinite if there is a fixed point $(F_{*},p_{*})$
on the first branch of $J/v(F)$. In such a case, we may have wave trains only
if $p>p_{*}$.
\end{itemize}
The previous construction of the wave train gives its velocity and period
$T\sim T_{p}+T_{t}$ as functions of the parameter $p$. The spatial period
$L$ is the integer part of $cT/\delta$. The number of QWs in the peak
(resp. the tail) region is the integer part of $c_{-}(p)T_{p}/\delta$
(resp. $c_{-}(p)T_{t}/\delta$). Our construction clearly fails if the number
of QWs in the peak region is smaller than one, i.e., if $c_{-}(p)T_{p}<\delta$.
Thus, $\delta$ has to be larger than $c_{-}(p)T_{p}$.

\section{Pulses moving upstream in a dc-current-biased photoexcited SL that
behaves as an excitable medium}
\label{sec:upstream}
\begin{figure}[!t]
\centerline{\includegraphics[width=0.5\linewidth]{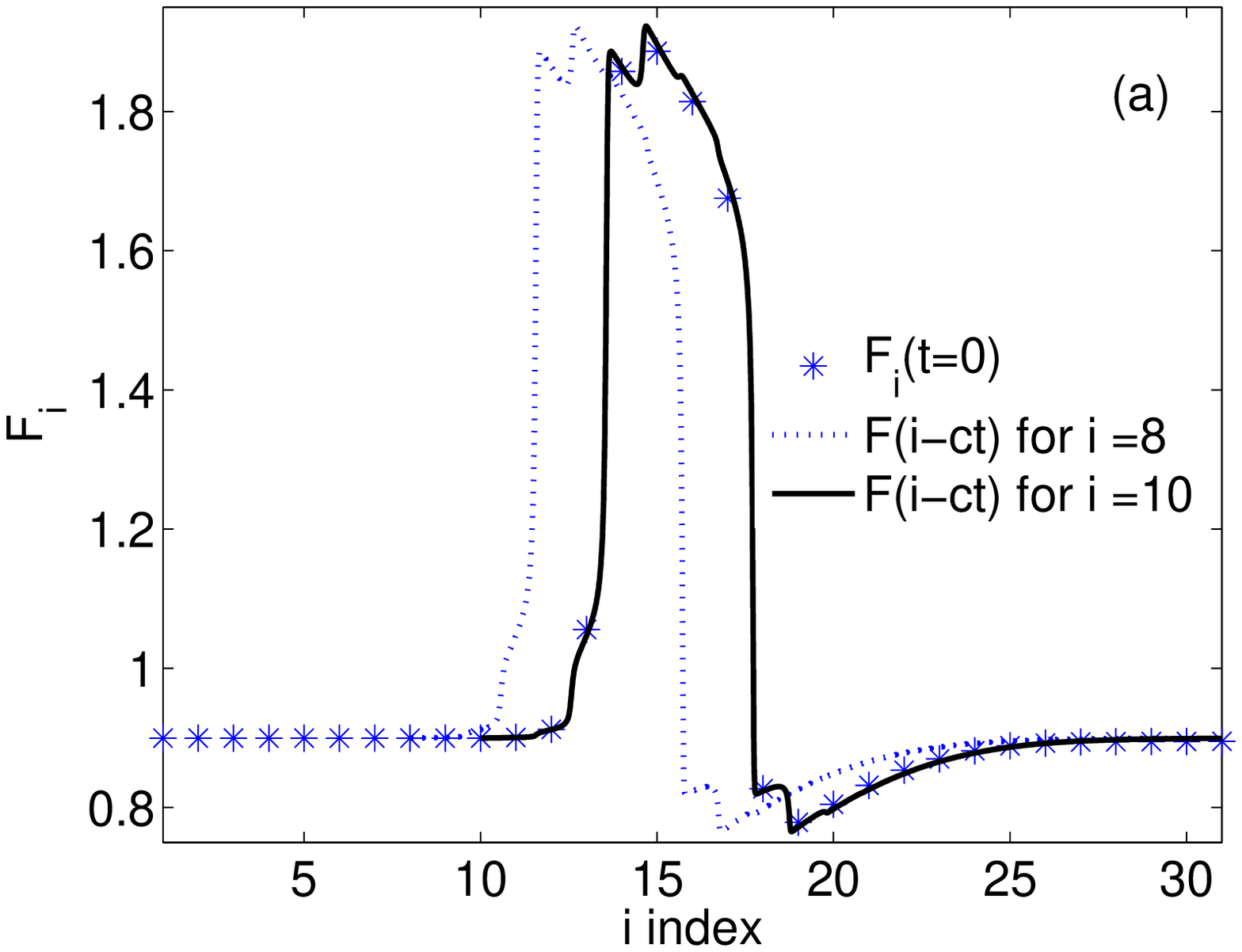}}
\centerline{\includegraphics[width=0.5\linewidth]{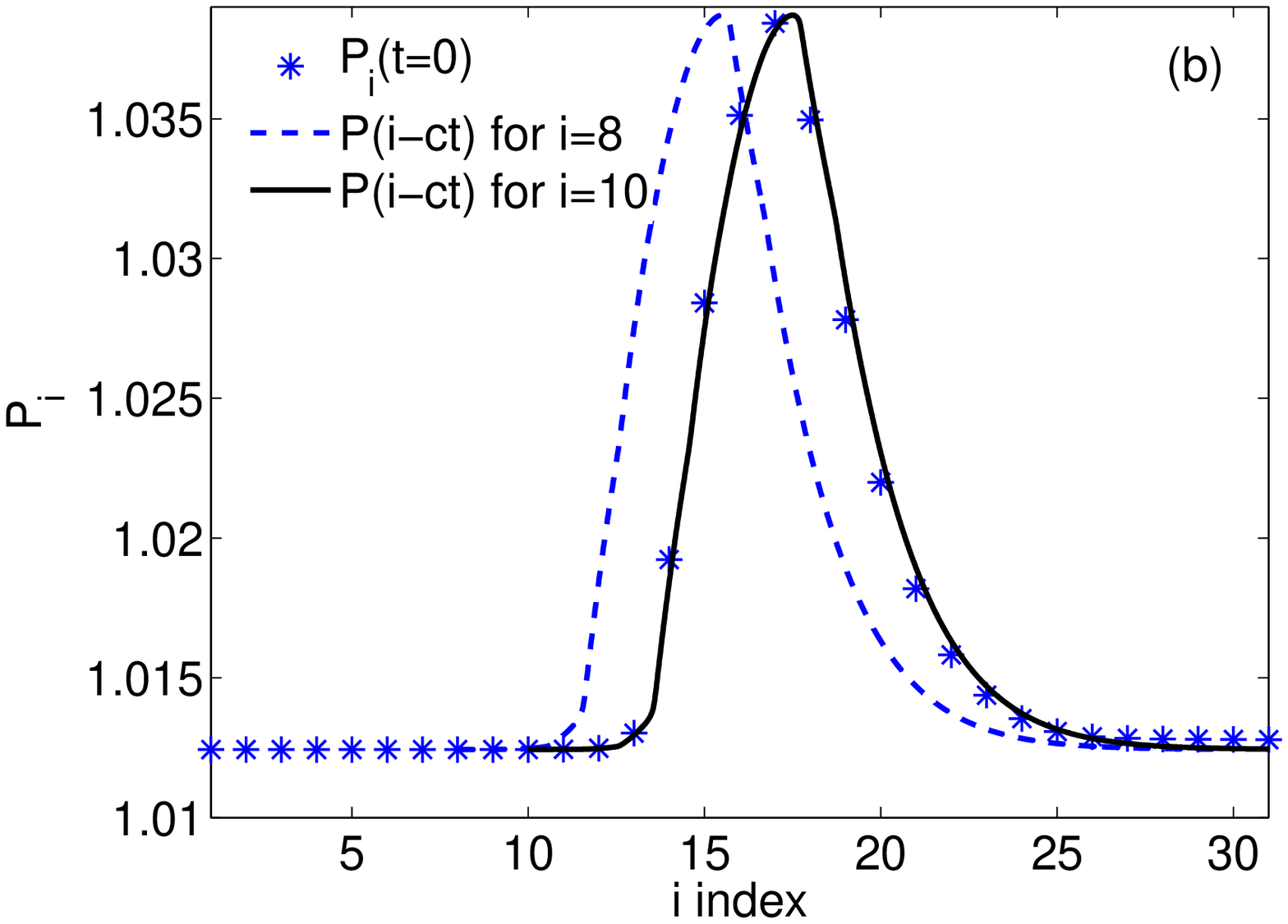}}
\caption{(Color online) (a) Numerically obtained field profile of a pulse moving with negative velocity for $J=1.009$, $\nu=70.5$, $\delta=4.05\times 10^{-4}$.
(b) Numerically obtained hole density profile of the same pulse.
The laser intensity is 479.7 kW/cm$^2$.}
\label{fig9}
\end{figure}

Numerical simulations of the complete model show pulses moving upstream
with a negative velocity, both under dc current bias and under dc voltage
bias. Although the field profile of these pulses is quite similar to that
of downstream moving pulses (cf. Figs.~\ref{fig6} and \ref{fig9}), there
are fundamental differences between them. The velocity of these pulses is
much smaller than that of downstream moving pulses, and they cannot be
approximated by one IF and one DF plus regions of slow variation of the
electric field. In fact, Fig.~\ref{fig4}(b) shows that, contrary to the
case of wave fronts moving with positive velocity depicted in
Fig.~\ref{fig4}(a), it is not possible for a DF and an IF to
move with the same negative velocity at a fixed $J$. Thus, our construction
of pulses in Section \ref{sec:pulses} cannot describe pulses that move
rigidly upstream with negative velocity.

\begin{figure}[!t]
\centerline{\includegraphics[width=0.5\linewidth]{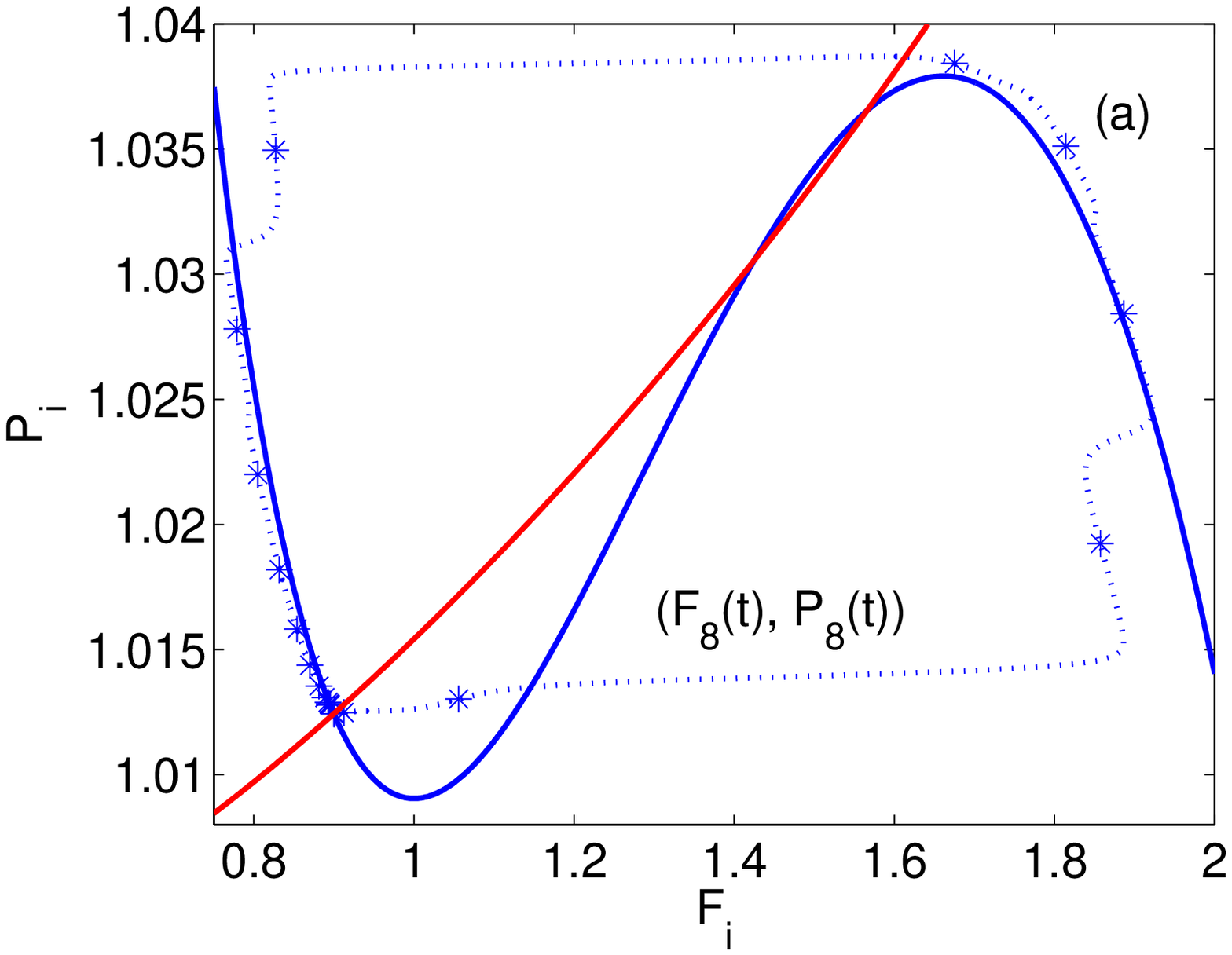}}
\centerline{\includegraphics[width=0.5\linewidth]{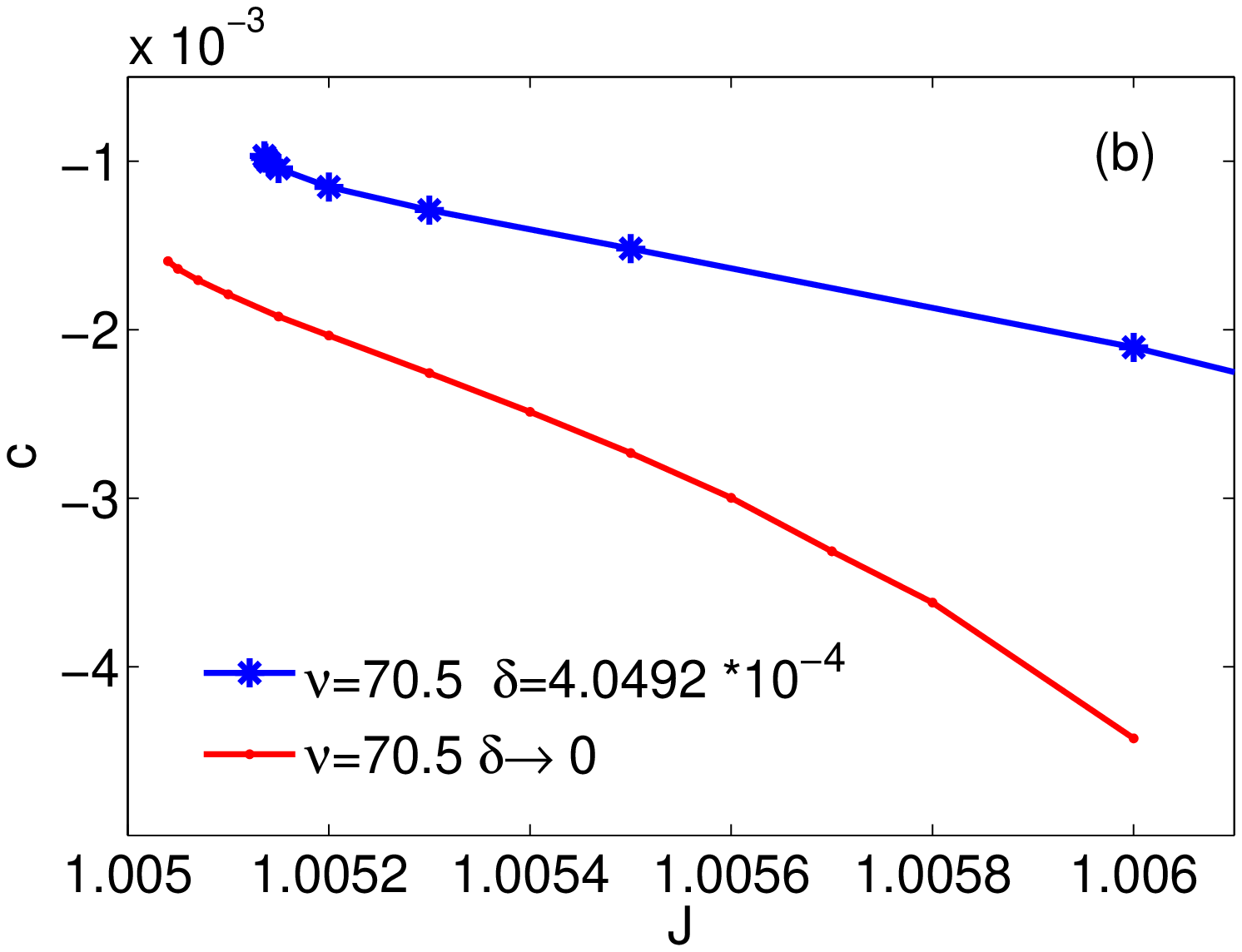}}
\caption{(Color online) (a) Phase plane showing the nullclines and the motion of the $8^{th}$ QW as the pulse traverses it for the profiles depicted in Fig.~\ref{fig9}.
(b) Comparison of the pulse velocity obtained from leading order perturbation
theory and that obtained from numerical simulations.}
\label{fig10}
\end{figure}

The asymptotic construction of pulses moving rigidly with negative velocity
under dc current bias is necessarily different from the case of downstream
moving pulses. The rigid upstream motion of pulses is saltatory:
although the velocity of the pulse is constant ($c<0$), there are periods
in which the QWs move slowly on the pulse field profile separated by fast
transitions as shown in Fig.~\ref{fig9}(a). Figure~\ref{fig9}(b) shows that
the hole density changes smoothly even at the fast transitions in the field
profile. During the slow periods, the field and hole density at the QWs
evolve according to the following equations:
\begin{eqnarray}
&& \left(p_{i}+\frac{F_{i}-F_{i-1}}{\nu}\right)v(F_{i}) - \left(
p_{i+1}-p_{i}+\frac{F_{i+1}+F_{i-1}-2F_{i}}{\nu}\right)D(F_{i})\nonumber\\
&& = J\;,
\label{eq38}\\
&&\frac{dp_{i}}{dt}=1 - r(F_{i})\, p_{i} \left(p_{i}+\frac{F_{i}-F_{i-1}}{\nu}
\right)\;,
\label{eq39}
\end{eqnarray}
which have been obtained by setting $\delta=0$ in Eqs.~(\ref{eq7})--(\ref{eq8}).
We have to solve Eq.~(\ref{eq38}) for the profile $\{F_{i}\}$ in terms of the
instantaneous values of the $\{p_{i}\}$ and insert the result in Eq.~(\ref{eq39}).
It turns out that there are several possible solutions corresponding to
integer shifts of the pulse profile $i\to i + m$, $m=0,\pm 1,\ldots$. The
implicit function theorem establishes that, starting from an appropriate
pulse-like initial condition, it is possible to find $F_{i}=F_{i}(\{p_{j}\})$,
provided the Jacobian determinant corresponding to Eq.~(\ref{eq38}) is not zero.
This condition holds until the QWs reach the points of abrupt field change
in Figs.~\ref{fig9}(a) or \ref{fig10}(a) (phase plane) at times $t=t_{a}$.
Then the Jacobian vanishes, and the field values change according to
\begin{eqnarray}
\frac{dF_{i}}{d\tau}=J+ \left(p_{i+1}-p_{i}+\frac{F_{i+1}+F_{i-1}-2F_{i}}{\nu}
\right)D(F_{i})\nonumber\\
-\left(p_{i}+\frac{F_{i}-F_{i-1}}{\nu}\right)v(F_{i})\;,
\label{eq40}
\end{eqnarray}
with $\tau=(t-t_{a})/\delta$ and $\{p_{i}\}$ frozen at their values at
$t_{a}$ given by Eqs.~(\ref{eq38})--(\ref{eq39}). As explained before, as
$\tau\to -\infty$ the $\{F_{i}(\tau)\}$ tends  to the solution of Eq.~(\ref{eq38})
for $t=t_{a}$, and it tends to the same profile shifted one step to the left:
$i\to i-1$ as $\tau\to +\infty$. Then another slow stage follows.
Figure~\ref{fig10}(b) shows the velocity of the pulse. It is approximately
the inverse time one QW spends in the short interval of $p$ between the
abrupt long jump in $F$ from the first branch of $J/v(F)$ to a value below
that in the third branch of $J/v(F)$ and the abrupt short jump in $F$ to
the third branch of $J/v(F)$, which occurs for a somewhat higher value of
$p$ as shown in Fig.~\ref{fig10}(a).

\begin{figure}[!t]
\centerline{\includegraphics[width=0.5\linewidth]{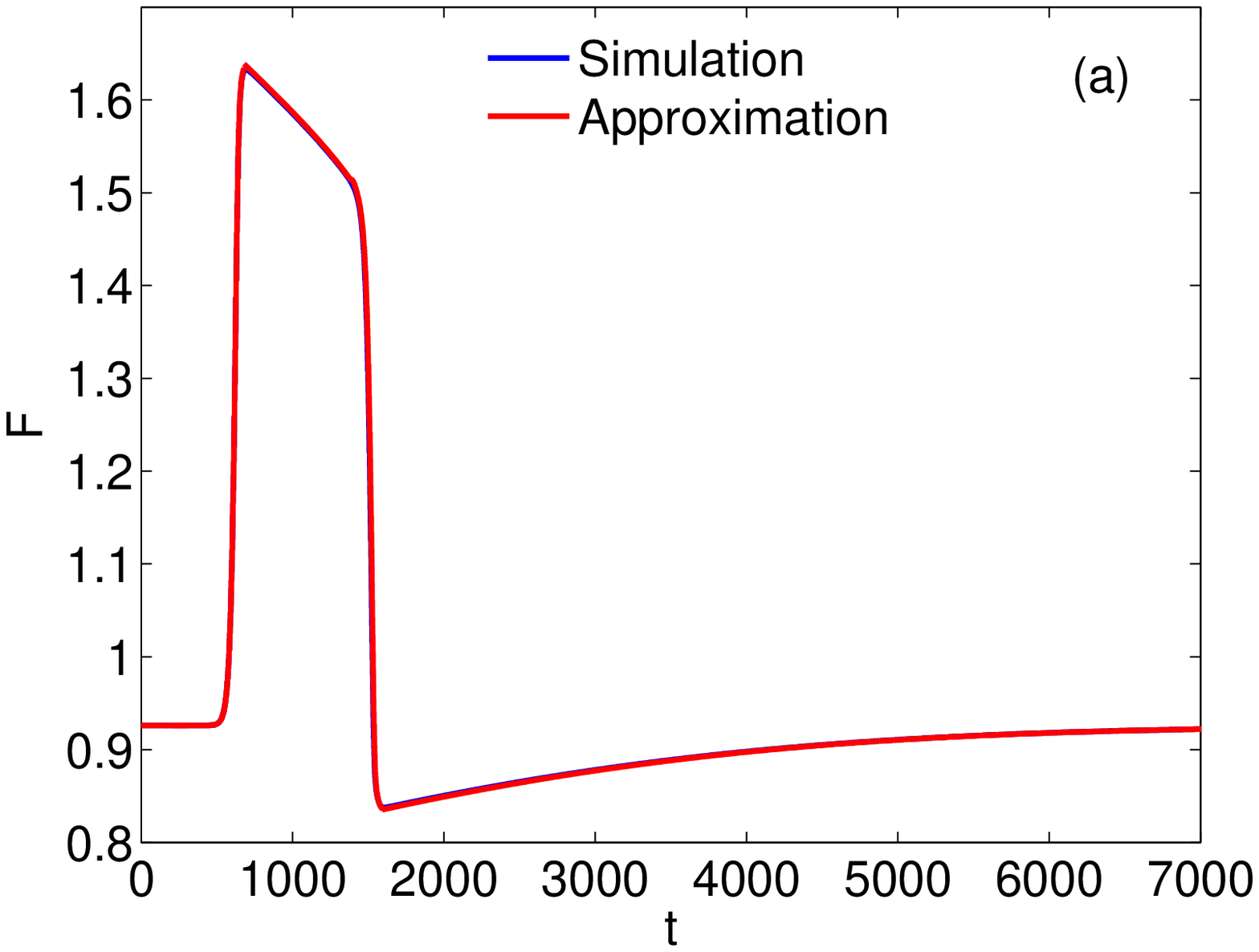}}
\centerline{\includegraphics[width=0.5\linewidth]{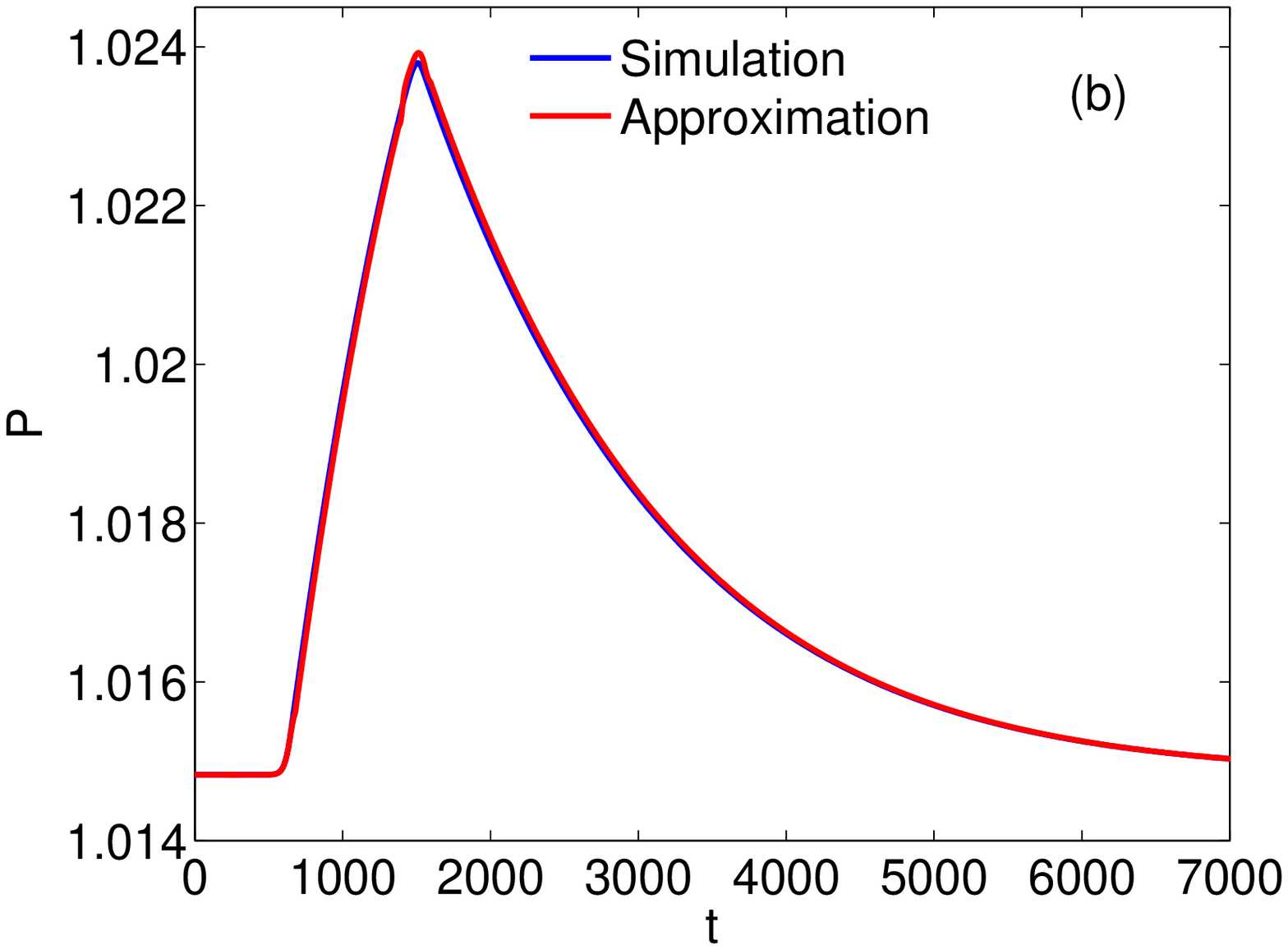}}
\caption{(Color online) (a) Reconstruction of a pulse of the electric field moving with a
positive velocity for $J=1.009$, $\nu=70.5$, $\delta=4.05\times 10^{-4}$ as
in Fig.~\ref{fig7}. (b) Reconstruction of the corresponding hole density
profile.}
\label{fig11}
\end{figure}

\begin{figure}[!t]
\centerline{\includegraphics[width=0.5\linewidth]{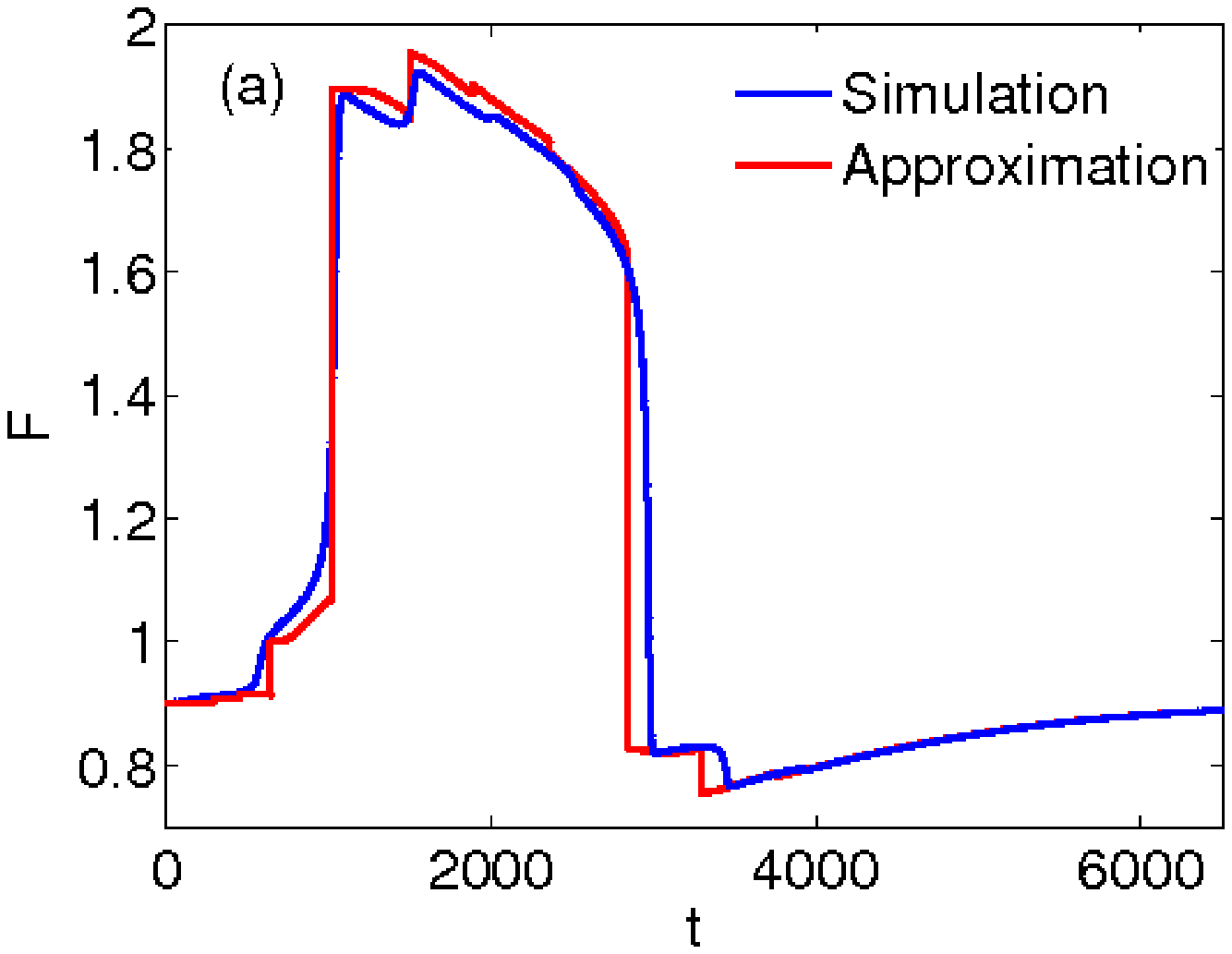}}
\centerline{\includegraphics[width=0.5\linewidth]{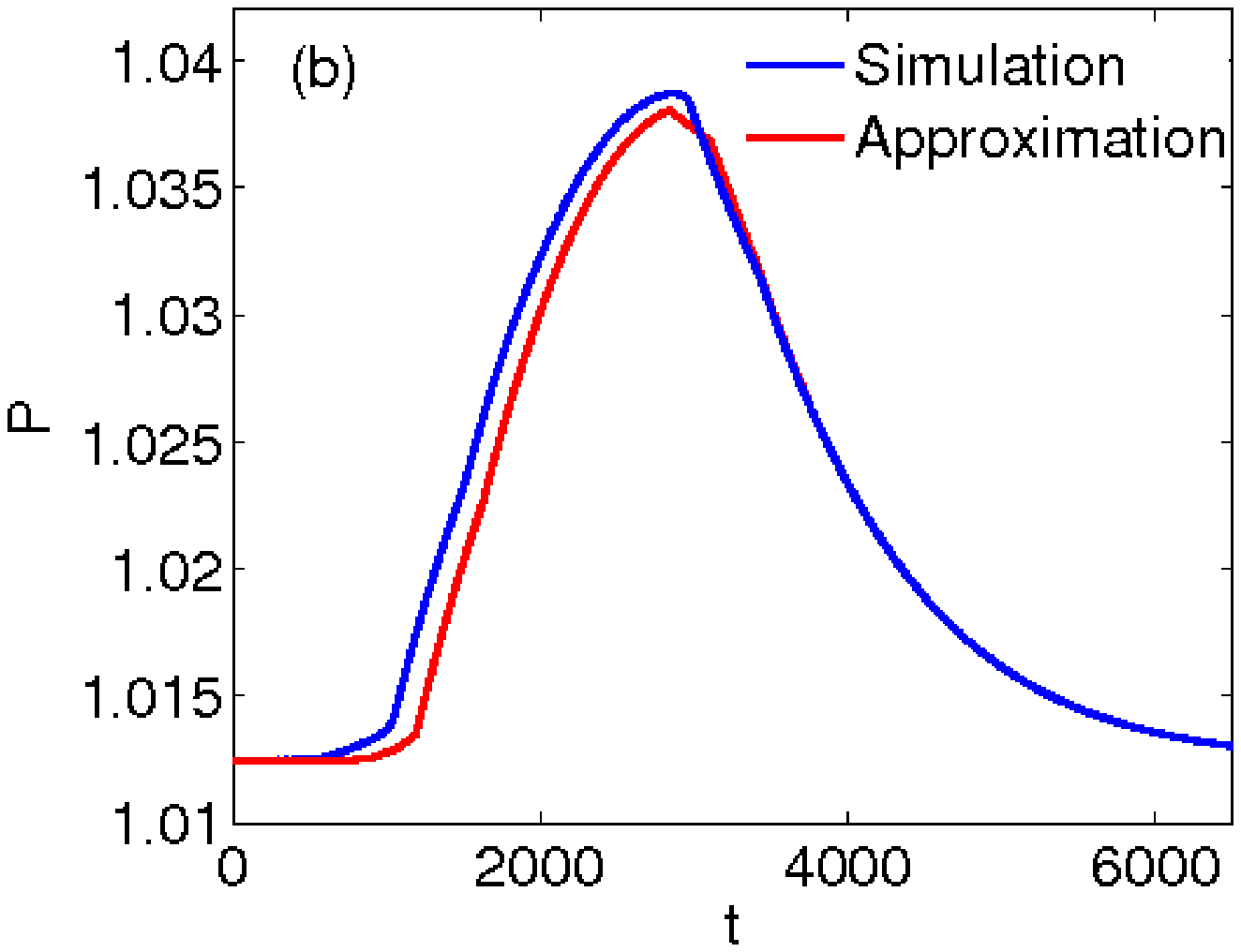}}
\caption{(Color online) (a) Reconstruction of a pulse of the electric field moving with a
negative velocity for $J=1.009$, $\nu=70.5$, $\delta=4.05\times 10^{-4}$. (b)
Reconstruction of the corresponding hole density profile.}
\label{fig12}
\end{figure}

Figures \ref{fig11}(a) and \ref{fig11}(b) compare the field and hole density
profiles of the reconstructed pulse (moving with positive velocity) and those
obtained by direct numerical solution of the equations. Similarly, Figures
\ref{fig12}(a) and \ref{fig12}(b) compare the field and hole density profiles
of the reconstructed pulse (moving with negative velocity) and those obtained
by direct numerical solution of the equations. The agreement is quite
reasonable and it could be improved by finding a corrected theory of the
pulse velocity.

 \section{Conclusions}
An undoped, strongly photoexcited, weakly coupled, dc-current-biased SL exhibits
a variety of pulse and wave front solutions. They can be constructed using
matched asymptotic expansions that exploit the large separation of time scales
(measured by the dimensionless parameter $\delta$) between the carrier dynamics
and the evolution of the electric field. On the time scale of the electron and
hole densities, the field follows adiabatically the profiles thereof.

For large photoexcitation, the field profile of a pulse typically consists of
slowly varying regions separated by sharp wave fronts. As in the case of the
FitzHugh-Nagumo model of nerve conduction, an asymptotic reconstruction of the
downstream-moving pulse to leading order in $\delta$ determines its velocity
as the value for which the velocities of the leading and trailing fronts are
the same. The comparison of this approximate velocity to that obtained by
direct solution of the full model equations is only fair. For a better agreement,
we have improved the description of the wave fronts comprising the pulse by
including order $\delta$ corrections. The resulting corrected theory describes
the pulses and their velocities much better. Pulses moving upstream the
electron flow exhibit saltatory motion: they alternate periods of very small
velocity with fast motion during short time intervals. We have reconstructed
them using matched asymptotic expansions in the limit of small $\delta$,
but the resulting theory is quite different from that of downstream-moving pulses.

\ack
The work was financially supported in part by the Spanish Ministry
of Science and Innovation under grant FIS2008-04921-C02-01.

\end{document}